\documentclass[reprint,aps,twocolumn,prd]{revtex4}

\usepackage{amssymb}
\usepackage{amsfonts}
\usepackage{graphicx}
\usepackage{epstopdf}
\usepackage{dcolumn}
\usepackage{amsmath}
\usepackage{latexsym,bm}
\usepackage{amsthm}
\usepackage{slashed}
\usepackage{float}
\usepackage{color}
\usepackage{url}
\usepackage{longtable}

\usepackage{soul}

\newcommand{\jht}[1]{{}}
\newcommand{\jt}[1]{{}}

\definecolor{cobalt}{RGB}{44, 98, 120}
\definecolor{celadon}{rgb}{0.67, 0.88, 0.69}
\definecolor{dm}{cmyk}{.20, 0, .30, 0}
\definecolor{burgundy}{rgb}{0.5, 0.0, 0.13}
\definecolor{plotBlue}{RGB}{94, 130, 181}

\newcommand \xoverline[2][0.75]{
    \sbox{\myboxA}{$\m@th#2$}
    \setbox\myboxB\null
    \ht\myboxB=\ht\myboxA
    \dp\myboxB=\dp\myboxA
    \wd\myboxB=#1\wd\myboxA
    \sbox\myboxB{$\m@th\overline{\copy\myboxB}$}
    \setlength\mylenA{\the\wd\myboxA}
    \addtolength\mylenA{-\the\wd\myboxB}
    \ifdim\wd\myboxB<\wd\myboxA
       \rlap{\hskip 0.5\mylenA\usebox\myboxB}{\usebox\myboxA}%
    \else
        \hskip -0.5\mylenA\rlap{\usebox\myboxA}{\hskip 0.5\mylenA\usebox\myboxB}%
    \fi}
\makeatother

\def \be {\begin{equation}}
\def \ee {\end{equation}}
\def \bsp {\begin{split}}
\def \esp {\end{split}}
\def \bea {\begin{eqnarray}}
\def \eea {\end{eqnarray}}
\def \nn {\nonumber}
\def\mc{\mathcal}

\def\mb{\mathbb}
\def\mbf{\mathbf}
\def \bp{\begin{pmatrix}}
\def\ep{\end{pmatrix}}

\def\ord{\mathrm{ord}}
\def\max{\mathrm{max}}
\def\min{\mathrm{min}}
\def\dim{\mathrm{dim}}

\begin{document}

\title{E-string spectrum and typical F-theory geometry}

\author{Jiahua Tian}
\affiliation{Department of Physics, Northeastern University, 
Boston, MA 02115, U.S.A.}
\email{tian.jiah@husky.neu.edu}
\author{Yi-Nan Wang}
\affiliation{Mathematical Institute, University of Oxford, 
Woodstock Road, Oxford, OX2 6GG, United Kingdom}
\email{yinan.wang@maths.ox.ac.uk}

\date{\today}

\begin{abstract}
In recent scans of 4D F-theory geometric models, it was shown that a dominant majority of the base geometries only support SU(2), $G_2$, $F_4$ and $E_8$ gauge groups. Moreover, most of these gauge groups are shown to couple to strongly coupled ``conformal matter'' sectors. For example, the $E_8$ gauge group can couple to the compactification of 6D E-string theory on a complex curve. In this paper, we initiate the investigation of these strongly coupled sectors by studying the spectrum of 6D E-string theory. We construct a resolved elliptic Calabi-Yau threefold of a non-minimal Weierstrass model, which contains a non-flat fiber with the topology of generalized del Pezzo surface. The spectrum of E-string theory then arises from M2 brane wrapping modes on various 2-cycles on the non-flat fiber. Finally, we discuss the compactification of these fields to 4D.
\end{abstract}

\keywords{}

\maketitle

\section{Introduction}

Over the past thirty years, people have gradually realized that string theory admits a huge number of vacuum solutions, forming the vast ``landscape''~\cite{Lerche:1986cx,Vafa:2005ui}. There are then two roads towards a string theory construction of our real world physics: the \textit{bottom-up} approach and the \textit{top-down} approach. In the bottom-up approach, one engineers a string theory geometric solution to make it contain our particle physics standard model. However, there is not a good reason to choose a specific initial condition in the first place. While in the alternative top-down approach, one attempts to classify all the possible string theory geometric solutions and study their statistical properties. If most of these string theory solutions share a common feature, we may say that this feature is a ``prediction'' from string theory. Moreover, we can choose a ``typical'' string theory geometry with all the common features, which serves as a natural starting point for model building.

Following the top-down approach, we choose the F-theory framework as our starting point, which is a powerful geometric description of strongly coupled IIB superstring theory~\cite{Vafa:1996xn,Morrison:1996na,Morrison:1996pp}. After compactifying F-theory on an elliptically fibered Calabi-Yau fourfold $X$, one obtains a 4D $\mc{N}=1$ supergravity theory. The F-theory ensemble contains some of the largest finite numbers of vacuum solutions in any string theory literature. For example, it was shown that there exists at least $\sim 10^{755}$ topologically distinct toric base threefolds that can support a 4D F-theory solution \cite{Halverson:2017ffz}, and this number is further enlarged to $10^{3,000}$ with a statistical approach~\cite{Taylor:2017yqr}. Moreover, on a single elliptic Calabi-Yau fourfold $\mc{M}_{\rm max}$ with base $B_{\rm max}$, there exists $\sim 10^{224,000}$ distinct choices of self-dual 4-form flux~\cite{Taylor:2015xtz}, each of which may give rise to a different vacuum solution. 
 
There is a stunning universal feature of the dominant majority of these 4D vacua: the geometric gauge groups are mostly in the form of 
\be
G=SU(2)^a\times G_2^b\times F_4^c\times E_8^d.\label{genericG}
\ee 
For example, the geometric gauge group of the 4D F-theory model on $\mc{M}_{\rm max}$ is $G=E_8^9\times F_4^8\times (G_2\times SU(2))^{16}$ \cite{Taylor:2015xtz}. In the combinatoric ensemble of $\frac{4}{3}\times 2.96\times 10^{755}$ toric base threefolds \cite{Halverson:2017ffz}, it was shown that $99.9995\%$ of toric threefolds  contain $E_8^{10}\times F_4^{18}$ as a subgroup of the geometric non-Higgsable gauge group. In \cite{Taylor:2017yqr}, there is an empirical formula on the number of simple non-Higgsable gauge group components on a base threefold $B$ with $h^{1,1}(B)\gtrsim 1,000$:
\begin{align*}
&N_{SU(2)}\cong\left[\frac{h^{1,1}(B)+1}{6}\right]\ ,\ N_{G_2}\cong\left[\frac{h^{1,1}(B)+1}{9}\right]\ , \\ &N_{F_4}\cong\left[\frac{h^{1,1}(B)+1}{24}\right]\ ,\ N_{E_8}\cong\left[\frac{h^{1,1}(B)}{68}\right].\label{Ngaugegroups}
\end{align*}

Another stunning result from the Monte Carlo approach~\cite{Taylor:2017yqr} shows that the majority of elliptic fibrations are non-minimal, or non-flat. Physically, this means that the 4D physical model is actually a supergravity coupled to modes from a 4D SCFT or a compactification of 6D SCFT on a complex curve. These modes are generally called ``conformal matter'' in our context~\cite{DelZotto:2014hpa,Apruzzi:2018oge}. For example, in~\cite{Taylor:2017yqr}, it was found that there are only $\sim 10^{250}$ ``good bases'' which can support a 4D Lagrangian supergravity theory\footnote{The generic elliptic fibration on many good bases has the issues of codimension-three non-minimal singularities~\cite{Candelas:2000nc,Braun:2013nqa,Achmed-Zade:2018idx} and terminal singularities~\cite{Arras:2016evy}, which have not been well understood. We assume that these geometric features will not invalidate the 4D Lagrangian description.}, among the  set of $\sim 10^{3,000}$ ``resolvable bases'' containing conformal matter sectors. 

It is then natural to ask:

\textit{Is it possible to embed the particle physics standard model into a 4D F-theory model on a typical base?}

If the answer to the above question is No, then there are two possible conclusions:
\begin{enumerate}
\item{It is extremely unnatural to find our real world in the F-theory landscape.}
\item{The typical geometries studied in \cite{Taylor:2015xtz,Taylor:2017yqr,Halverson:2017ffz} are disfavoured by some unknown physical dynamics.}
\end{enumerate}

Due to the absence of conventional GUT gauge groups: SU(5), SO(10) and $E_6$ on these geometries, the most natural way of embedding standard model seems to be the inclusion of $SU(3)\times SU(2)\times U(1)$ into a single $E_8$. However, if one only utilizes the ``bulk fields'' on a stack of $E_8$ 7-branes, one faces the problem of no net chirality~\cite{Tatar:2006dc} and the absence of Yukawa coupling terms~\cite{Beasley:2008dc}. The only possible realization of 4D chiral matter is then using localized matter fields on a complex curve $\Sigma\subset S$, which are charged under $E_8$. However, in F-theory, the only possible charged matter fields under $E_8$ are the strongly coupled matter fields! For example, we can consider a 6D rank-1 E-string theory compactified on the curve $\Sigma$. 

The E-string theory is a strongly coupled 6D (1,0) SCFT with $E_8$ global symmetry, which is the world-volume theory of M5 branes ending on a M9 brane \cite{Ganor:1996mu,Seiberg:1996vs}. The fundamental objects of the E-string theory are self-dual tensionless strings, which carry an infinite tower of massless higher spin modes. In this paper, we study the particle spectrum of rank-1 E-string theory from M/F-theory duality. We start with a generic elliptic fibration $X_3$ over the Hirzebruch surface $\mb{F}_{11}$, which has a non-minimal $E_8$ non-Higgsable gauge group on the $(-11)-$curve. The resolution of $X_3$ is a smooth Calabi-Yau threefold $\hat{X}_3$, which has a non-flat fiber $S_{nf}$ with the topology of a generalized del Pezzo surface $gdP_2$. In the M-theory dual picture, the M2 brane wrapping modes on the complex curves $C\subset S_{nf}$ will give rise to an infinite massless higher spin tower in the 6D F-theory limit. We find that the lowest spin modes are a 6D hypermultiplet in $\mbf{248}$ representation of $E_8$, a vector multiplet in representation $\mbf{3875}$ and a Rarita-Schwinger multiplet in representation $\mbf{147250}$. 

After the 6D spectrum is obtained, we study the reduction of these fields on a complex curve $\Sigma$ with the inclusion of gauge flux. In order to preserve 4D $\mc{N}=1$ supersymmetry, one needs to introduce topological twist to organize the fields into various $\mc{N}=1$ supermultiplets. Nonetheless, it is unclear which 4D fields will remain massless after the coupling with gravity and gauge fields.

The structure of this paper is as follows: in section~\ref{sec:F-theory}, we review the basic setups of F-theory and show the common characteristics of a typical 4D F-theory geometry. In section~\ref{sec:spectrum}, we study the spectrum of the 6D E-string theory via the M-theory resolution picture. In section~\ref{sec:4D}, we discuss the compactification of 6D E-string spectrum to 4D. Finally, we summarize the paper with some discussions in section~\ref{sec:discussion}. 

\section{Characteristics of a typical 4D F-theory geometric model}
\label{sec:F-theory}

\subsection{Basics of F-theory}

A good introduction to F-theory can be found in~\cite{Weigand:2010wm,Weigand:2018rez}, and we will only review the parts that are necessary for our purpose. In our setup, we consider F-theory compactified on an elliptically fibered Calabi-Yau space $X_{d+1}$ over a complex base $B_d$ with $d$-complex dimensions, which leads to a physical model in Minkowski space-time $\mb{R}^{9-2d,1}$. The elliptic fibration is described by a Weierstrass model:
\be
y^2=x^3+fx+g,\label{Weierstrass}
\ee
where $f\in\mc{O}(-4K_B)$ and $g\in\mc{O}(-6K_B)$ are sections of the line bundles $-4K_B$ and $-6K_B$ on $B$. The Weierstrass model (\ref{Weierstrass}) is singular at the discriminant locus
\be
\Delta=4f^3+27g^2=0.
\ee
For a valid geometric model in F-theory, we require that there exists a crepant resolution $\rho:\hat{X}_{d+1}\rightarrow X_{d+1}$ of the singular space $X_{d+1}$, such that $\hat{X}_{d+1}$ is an elliptic Calabi-Yau manifold with only terminal singularities. In the M-theory dual picture compactified on $\hat{X}_{d+1}$, there exists singular fibers over certain subset $S\subset B_d$. M2 brane wrapping modes on these singular fibers will give rise to gauge fields and matter fields in the F-theory limit, where all the fiber directions are shrunk to zero size. 

Especially, when $S$ is complex codimension-one, or equivalently a divisor of $B$, the singular fiber over $S$ is a collection of exceptional $\mb{P}^1$s forming an affine Dynkin diagram of a Lie algebra $\mathfrak{g}$. The M2 branes wrapping these exceptional $\mb{P}^1$s will give rise to the W-bosons in the adjoint representation $\text{adj}(G)$, where the non-Abelian gauge group $G$ has the Lie algebra $\mathfrak{g}$. The Cartan generators of $\text{adj}(G)$ are reduced from the $C_3$ field in M-theory:
\be
C_3=\sum_i A\wedge\omega_i,
\ee
where $\omega_i$ is the $(1,1)$-form Poincar\'{e} dual to the exceptional divisor $E_i$, which is a fibration of an exceptional $\mb{P}^1$ over $S$. 

In the IIB/F-theory picture, the non-Abelian gauge fields are the open string modes of a stack of 7-branes wrapping $S\times \mb{R}^{9-2d,1}$. The dictionary between the gauge group $G_S$ on this stack of 7-branes and the properties of $(f,g,\Delta)$ near $S$ is well-studied, see table~\ref{t:gaugegroup}. Besides the information of the order of vanishing of $(f,g,\Delta)$ on $S$, one also need to write down the ``monodromy cover polynomial''~\cite{Grassi:2011hq} for some of the Kodaira fiber types. For example, if we have ord$_S(f,g,\Delta)=(2,3,6)$, then we need to examine the polynomial $M(\psi)=\psi^3+f_2\psi+g_3$. If $M(\psi)$ can be reduced to a product of three polynomials in $\psi$, then $G_S=$SO(8). If $M(\psi)$ can only be reduced to a product of two polynomials in $\psi$, then $G_S=$SO(7). Otherwise, if $M(\psi)$ is completely irreducible, then $G_S=G_2$.

\begin{table*}
\caption{\label{t:gaugegroup} The list of the criteria for all the gauge groups in F-theory on a divisor $S:u=0$. The coefficients $f_i$ and $g_i$ are from the expansion $f=\sum_i f_i u^i$ and $g=\sum_i g_i u^i$ near $u=0$. The fourth column is the Kodaira type of the singlar fiber, and $M(\psi)$ is the monodromy cover polynomial. When $M(\psi)$ is completely irreducible, the gauge group is given by the leftmost one. When $M(\psi)$ is completely reducible, the gauge group is given by the rightmost one.}
\begin{ruledtabular}
\begin{tabular}{|c|c|c|c|c|c|}
ord$(f)$ & ord$(g)$ & ord$(\Delta)$ &  & $M(\psi)$ & Gauge group $G_S$\\
\hline
0 & 0 & 2 & $I_2$ & - & SU(2)\\
0 & 0 & $n\geq 3$ & $I_n$ & $\psi^2+(9g/2f)|_{u=0}$ & Sp$\lfloor\frac{n}{2}\rfloor$ or SU$(n)$\\
1 & $\geq 2$ & 3 & $III$ & - & SU(2)\\
$\geq 2$ & 2 & 4 & $IV$ & $\psi^2-g_2$ & SU(2) or SU(3)\\
$\geq2$ & $\geq 3$ & 6 & $I_0^*$ & $\psi^3+f_2\psi+g_3$ & $G_2$ or SO(7) or SO(8)\\
2 & 3 & $2n+1(n\geq 3)$ & $I_{2n-5}^*$ & $\psi^2+\frac{1}{4}\Delta_{2n+1}(2uf/9g)^3|_{u=0}$ & SO$(4n-3)$ or SO$(4n-2)$\\
2 & 3 & $2n+2(n\geq 3)$ & $I_{2n-4}^*$ & $\psi^2+\Delta_{2n+2}(2uf/9g)^2|_{u=0}$ & SO$(4n-1)$ or SO$(4n)$\\
$\geq 3$ & 4 & 8 & $IV^*$ & $\psi^2-g_2$ & $F_4$ or $E_6$\\
3 & $\geq 5$ & 9 & $III^*$ & - & $E_7$\\
$\geq 4$ & 5 & 10 & $II^*$ & - & $E_8$\\
\end{tabular}
\end{ruledtabular}
\end{table*}

For the other cases not included in table~\ref{t:gaugegroup}, i. e. $\ord_S(f,g,\Delta)\geq(4,6,12)$ on a divisor $S$, there does not exist a crepant resolution of (\ref{Weierstrass}). For this reason, such a Weierstrass model cannot describe a supersymmetric vacuum solution in $\mb{R}^{9-2d,1}$, and we will not accept them in any circumstances. Similar problem arises when $\ord_C(f,g,\Delta)\geq(8,12,24)$ on a codimension-two locus $C$, or $\ord_P(f,g,\Delta)\geq(12,18,36)$ on a codimension-three locus $P$, and so on.

The story of singular fibers over a codimension-two locus $C\subset B_d$ is more intricate, and there does not yet exist a complete classification. Especially, when the order of vanishing of $(f,g)$ on $C$ satisfies
\be
\ord_C(f)\geq 4\ ,\ \ord_C(g)\geq 6,
\ee
there will be non-flat fiber components in the resolved space $\hat{X}_{d+1}$. A non-flat fiber $S_{nf}$ is a submanifold of $\hat{X}_{d+1}$ with complex dimension $\text{dim}(S_{nf})>1$, which is entirely in the fiber direction. We will provide detailed examples of non-flat fibrations in section~\ref{sec:resolution}.

In the discussion of elliptic fibration over a specific base manifold $B_d$, we often use the notion of non-Higgsable phase, where $f$ and $g$ are chosen to be generic sections of line bundles $-4K_B$ and $-6K_B$ respectively. In this phase, $X_{d+1}$ is a generic elliptic fibration over $B_d$, and the geometric gauge groups $G$ are minimal among all the possible fibrations over $B_d$, which are called non-Higgsable gauge groups~\cite{Morrison:2012np,Morrison:2014lca}. The only possible non-Higgsable gauge groups are SU(2), SU(3), $G_2$, SO(7), SO(8), $F_4$, $E_6$, $E_7$ and $E_8$.

If ord$_S(f,g,\Delta)\geq(4,6,12)$ on a divisor $S$ for the generic fibration over $B_d$, then none of the elliptic fibrations over $B_d$ can be crepantly resolved. Such a base manifold $B_d$ cannot support any elliptic Calabi-Yau manifold, and it is not allowed in the landscape of F-theory geometric solutions. For example, an elliptic fibration over the Hirzebruch surface $\mb{F}_n\equiv \mb{P}(\mc{O}(0)+\mc{O}(-n))$ always has codimension-one (4,6) singularity on its rational $(-n)$-curve for any $n\geq 13$. 

On the other hand, if $\ord_S(f,g,\Delta)\geq(4,6,12)$ on a codimension-two locus $C$ or $\ord_P(f,g,\Delta)\geq(8,12,24)$ on a codimension-three locus $P$  for the generic fibration over $B_d$, then any elliptic fibration over $B_d$ has the property of being a non-flat fibration. We typically call these locus codimension-two (4,6) or codimension-three (8,12) locus. This type of the base manifold $B_d$ is called a ``resolvable base'' if it can be blown up a finite number of times and transformed into a  ``good base'' without any codimension-two (4,6,12) or codimension-three (8,12,24) locus~\cite{Taylor:2017yqr}. 

\subsection{Ensemble of toric base threefolds}
\label{sec:ensemble}

A complete classification of the geometric vacuum solutions in 4D F-theory involves the following four steps:

\begin{enumerate}
	\item Classify all the topologically distinct base threefolds $B_3$.
	\item For each $B$, classify all the different elliptic fibrations $X_4$ over $B_3$.
	\item For each elliptic Calabi-Yau fourfold $X_4$, classify other discrete data such as the $G_4$ flux~\cite{Douglas:2003um,Denef:2004ze,Ashok:2003gk}.
	\item Study more detailed physics such as the moduli stabilization~\cite{Denef:2005mm,Cota:2017aal}, SUSY breaking and the mass of massive particles. 
\end{enumerate}

Even at the first step, the set of base threefolds is huge and there is no estimation on its total number. Nonetheless, the subset of the toric base threefolds has been explored in~\cite{Halverson:2015jua,Taylor:2015ppa, Halverson:2016tve, Halverson:2017ffz,Taylor:2017yqr}. The relatively most complete picture of the set of toric resolvable bases is achieved via a Monte Carlo random blow up approach~\cite{Taylor:2017yqr}. 

In the following discussion, we label the toric rays of a toric threefold base $B_3$ by $v_1,v_2,\cdots,v_n$, and their local hypersurface equations by $z_1=0$, $z_2=0$, $\dots$, $z_n=0$. The holomorphic monomials $m_f(u)$ and $m_g(u)$ in the Weierstrass polynomials $f$ and $g$ correspond to the lattice points in the Newton polytopes $\mc{F}$ and $\mc{G}$:
\be
\mc{F}=\{u\in\mb{Z}^3|\forall v_i\ ,\ \langle u,v_i\rangle\geq -4\}\ ,\ m_f(u)=\prod_{i=1}^n z_i^{\langle u,v_i\rangle+4},
\ee
\be
\mc{G}=\{u\in\mb{Z}^3|\forall v_i\ ,\ \langle u,v_i\rangle\geq -6\}\ ,\ m_g(u)=\prod_{i=1}^n z_i^{\langle u,v_i\rangle+6}.
\ee
Then the orders of vanishing of $f$ and $g$ on a toric divisor $D_i:z_i=0$ are 
\be
\bsp
&\text{ord}_{D_i}(f)=\min(\langle u,v_i\rangle+4)|_{u\in\mc{F}},\\
&\text{ord}_{D_i}(g)=\min(\langle u,v_i\rangle+6)|_{u\in\mc{G}},
\end{split}
\ee
and the order of vanishing of $f$ and $g$ on a toric curve $D_i D_j$ corresponding to a 2D cone $v_i v_j$ is
\be
\bsp
&\text{ord}_{D_i D_j}(f)=\min(\langle u,v_i\rangle+\langle u,v_j\rangle+8)|_{u\in\mc{F}},\\
&\text{ord}_{D_i D_j}(g)=\min(\langle u,v_i\rangle+\langle u,v_j\rangle+12)|_{u\in\mc{G}}.
\end{split}
\ee
The blow up of a toric curve $D_i D_j$ corresponds to an addition of a new ray $v_{n+1}=v_i+v_j$ and a subdivision of the 2D cone $v_i v_j$, and the blow up of a toric point $D_i D_j D_k$ corresponds to an addition of a new ray $v_{n+1}=v_i+v_j+v_k$ and a subdivision of the 3D cone $v_i v_j v_k$.

Starting with the base $a_1=\mb{P}^3$, we generate a random blow up sequence $s:a_1\rightarrow a_2\rightarrow \cdots\rightarrow a_N$, where each step is a blow up of a toric curve or a point, and each choice has the same probability. We require that all the bases $a_i$ in the sequence are resolvable by making sure that the origin $(0,0,0)$ is in the interior of $\mc{G}$~\cite{Taylor:2017yqr}. During this blow-up process, we record the number of possible blow up choices $N_{\rm up}(a_i)$ from a base $a_i$ and the number of possible blow down choices $N_{\rm down}(a_i)$ to another smooth resolvable base. Finally, we will hit an end point base $a_N$ where $N_{\rm up}(a_N)=0$.

\begin{figure}
\centering
\includegraphics[height=5cm]{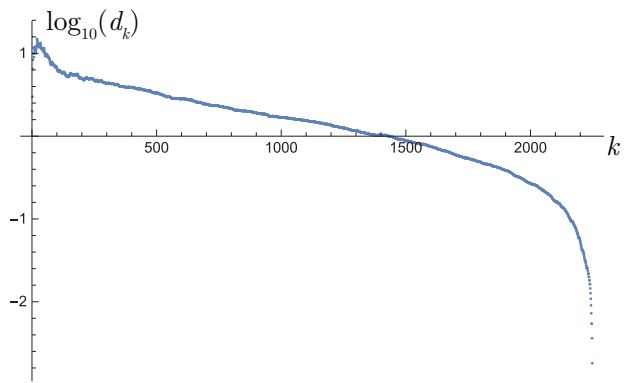}
\caption[E]{\footnotesize The factor $d_k=\frac{N_{\rm up}(a_k)}{N_{\rm down}(a_{k+1})}$ in terms of $h^{1,1}(B)=k$.}\label{f:logd}
\end{figure}

The total number of the toric resolvable bases with each $h^{1,1}(B)$ can be roughly estimated by a dynamic weight:
\be
D(h^{1,1}(B))=\prod_{i=1}^{h^{1,1}(B)-1}d_k\ ,\ d_k=\frac{N_{\rm up}(a_k)}{N_{\rm down}(a_{k+1})}.
\ee
For example, we plot the factors $d_k$ of a blow up sequence $s:a_1\rightarrow a_2\rightarrow \cdots\rightarrow a_N$ in figure~\ref{f:logd}. Hence one can see that the total number of the bases reaches the maximum around the ``turning point'' where $d_k$ becomes less than 1 for the first time.

After generating 2,000 different random blow up sequences, we take the average $D(h^{1,1}(B))$ to estimate the total number of bases for each $h^{1,1}(B)$:
\be
N(h^{1,1}(B))\approx \langle D(h^{1,1}(B))\rangle.
\ee
As a result, we estimated that there are $>10^{3,000}$ resolvable bases in~\cite{Taylor:2017yqr}. The good bases, which have no codimension-2 (4,6) locus, are almost entirely concentrated at the end points with a total number of $>10^{250}$. In the sampling of~\cite{Taylor:2017yqr}, it turns out that the non-Higgsable gauge groups on the end point bases are always in the form of
\be
G_{nH}=SU(2)^a\times G_2^b\times F_4^c\times E_8^d\times H,
\ee 
where $H=$SU(3), SO(8) for a small proportion of the base. The reason is that the number of monomials $|\mc{F}|$, $|\mc{G}|$ are extremely small for a generic base threefold with $h^{1,1}(B)\gtrsim 100$ generated in this random blow up sequence, see figure~4 of~\cite{Taylor:2017yqr}. Typically, there is only one monomial in $|\mc{F}|$, and the order of vanishing of $f$ on toric divisors is always 4. Comparing with table~\ref{t:gaugegroup}, one can see that the only possible geometric gauge groups (not only non-Higgsable gauge groups) under this condition are SU(2), SU(3), $G_2$, SO(8), $F_4$, $E_6$ and $E_8$. The gauge groups SU(2), $G_2$, $F_4$ and $E_8$ are far more common, since the other gauge groups require special reducibility conditions on the monodromy cover polynomial. Additionally, It is impossible to tune an SU(5) or SO(10) GUT gauge group on any of these toric divisors, since the order of vanishing of $f$ is already at least 4.

\subsection{A typical toric base threefold}
\label{sec:btp}

To study the property of a typical resolvable base in the sequence $s:a_1\rightarrow a_2\rightarrow \cdots\rightarrow a_N$, we define the turning point base $B_{tp}$ to be the base $a_i$ in the sequence with the smallest $i$ that satisfies
\be
N_{up}(a_i)\leq N_{down}(a_i).
\ee
From the discussions in section~\ref{sec:ensemble}, we can see that $B_{tp}$ represents the most common base at least in this single sequence $s$. As an example, there is a turning point base $B_{tp}$ with $h^{1,1}(B_{tp})=1385$ and the non-Higgsable gauge group
\be
G_{nH}=E_8^{34}\times F_4^{82}\times G_2^{192}\times SU(2)^{260}.
\ee

All of the 34 $E_8$ divisors on this $B_{tp}$ have codimension-two (4,6) locus on them (or simply called (4,6) curve). Denote a divisor with $E_8$ by $S:u=0$, and the toric divisors intersecting $S$ are $D_i(i=1,\dots,n):z_i=0$, then the local Weierstrass model near $S$ can be expanded as:
\be
y^2=x^3+f_4 u^4 x+g_5 u^5+\mc{O}(u^6).
\ee
We further rewrite $g_5$ as
\be
g_5=\tilde{g}_5 u^5\left(\prod_{i=1}^n z_i^{a_i}\right),
\ee
where $a_i$ is the order of vanishing of $g_5$ on divisor $D_i$.

If $a_i>0$, then the $\mb{P}^1$ curve $u=z_i=0$ is a (4,6) curve. Besides that, if $\tilde{g}_5$ is not a simple complex number, the curve $u=\tilde{g}_5=0$ is a (4,6) curve as well. To compute the genus of the curve $\tilde{g}_5=u=0$, notice that the hypersurface equation $\tilde{g}_5=0$ corresponds to the divisor class
\be
D(\tilde{g}_5)=-6K_B-5S-\sum_{i=1}^n a_i D_i.
\ee
Then we can compute the genus $\tilde{g}$ of the curve $u=\tilde{g}_5=0$ by the adjunction formula on $B_{tp}$:
\be
2\tilde{g}-2=(K_B+S+ D(\tilde{g}_5))\cdot D(\tilde{g}_5)\cdot S.
\ee

We classify the different types of (4,6) curves $u=z_i=0$ based on the gauge group on the divisor $z_i=0$. In our case, there are five different types: $E_8\times\varnothing$, $E_8\times$SU(2), $E_8\times G_2$, $E_8\times F_4$ and $E_8\times E_8$. The type of (4,6) curve $u=\tilde{g}_5=0$ is of $E_8\times\varnothing$ type if $\tilde{g}_5=0$ is irreducible. We list the total number of (4,6) curves with each type and $\tilde{g}$ for each divisor with $E_8$ gauge group on $B_{tp}$ in table~\ref{t:cod2}.

\begin{table}
\centering
\begin{tabular}{|c|c|c|c|c|c|}
\hline
\hline
$E_8\times\varnothing$ & $E_8\times$SU(2) & $E_8\times G_2$ & $E_8\times F_4$ & $E_8\times E_8$ & $\tilde{g}$\\
 \hline
2 & 4 & 2 & 1 & 0 & -\\
11 & 3 & 3 & 2 & 0 & -\\
11 & 5 & 7 & 2 & 0 & 1\\
2 & 4 & 2 & 1 & 0 & -\\
22 & 13 & 8 & 3 & 3 & -\\
1 & 2 & 1 & 0 & 0 & 0\\
5 & 1 & 1 & 0 & 0 & 0\\
3 & 1 & 0 & 1 & 0 & -\\
5 & 6 & 2 & 0 & 0 & -\\
4 & 0 & 2 & 1 & 0 & 0\\
0 & 3 & 0 & 0 & 0 & -\\
3 & 1 & 1 & 0 & 0 & -\\
1 & 2 & 0 & 0 & 0 & 0\\
0 & 2 & 0 & 2 & 0 & -\\
3 & 2 & 1 & 0 & 0 & 0\\
4 & 1 & 3 & 0 & 0 & 0\\
4 & 2 & 0 & 0 & 0 & -\\
5 & 3 & 1 & 3 & 1 & -\\
3 & 0 & 1 & 1 & 0 & 0\\
5 & 0 & 1 & 1 & 0 & 0\\
1 & 1 & 0 & 1 & 0 & 0\\
3 & 2 & 1 & 0 & 0 & 0\\
6 & 2 & 2 & 0 & 0 & -\\
1 & 2 & 0 & 1 & 0 & 0\\
0 & 1 & 1 & 1 & 0 & 0\\
0 & 4 & 0 & 0 & 1 & -\\
2 & 1 & 0 & 0 & 0 & 0\\
1 & 0 & 0 & 0 & 1 & 0\\
0 & 0 & 0 & 2 & 0 & -\\
2 & 2 & 1 & 3 & 1 & 0\\
3 & 0 & 0 & 0 & 0 & 0\\
0 & 0 & 1 & 0 & 0 & 0\\
0 & 0 & 0 & 1 & 0 & 0\\
0 & 0 & 0 & 0 & 0 & 0\\
\hline
\end{tabular}
\caption[E]{\footnotesize The number of (4,6) curves of each type on each $E_8$ divisor. The genus $\tilde{g}$ of the curve $\tilde{g}_5=u=0$ is also listed. The last entry of each row is a ``-'' if the $\tilde{g}_5$ is a non-zero complex number which simply means that the curve $\tilde{g}_5=u=0$ does not exist.}\label{t:cod2}
\end{table}

As one can see, there can be 22 $E_8\times\varnothing$ type, 13 $E_8\times$SU(2) type, 8 $E_8\times G_2$ type, 3 $E_8\times F_4$ type and $E_8\times E_8$ type  (4,6) curves on a single $E_8$ gauge group! Note that even for the last $E_8$ gauge group in table~\ref{t:cod2}, there is a $E_8\times\varnothing$ type (4,6) curve on a rational curve $\tilde{g}_5=u=0$. If we want to embed the standard model gauge group SU(3)$\times$SU(2)$\times$ U(1) into a single $E_8$ gauge group, it is crucial to understand the physics of these (4,6) curves. 

\subsection{The geometry with most flux vacua} 
\label{sec:max}

Before the discussion of the physics of (4,6) curves, we will briefly review the geometry of elliptic Calabi-Yau fourfold $\mc{M}_{\rm max}$ with the largest number of flux vacua~\cite{Taylor:2015xtz}. Since the total number of self-dual $G_4$ flux on $\mc{M}_{\rm max}$ has the order of $10^{227,000}$, which is overwhelmingly larger than the total number of base geometries we have estimated earlier, one should expect that $\mc{M}_{\rm max}$ is the most natural choice for an F-theory geometry if each flux choice is given the identical weight.

$\mc{M}_{\rm max}$ is a generic elliptic fibration over a toric threefold base $B_{\rm max}$. $B_{\rm max}$ is itself a $B_2$ bundle over $\mb{P}^1$, where $B_2$ is a compact toric surface with toric cyclic representation $(0,+6,-12//-11//-12//-12//-12//-12//-12//-12//-12)$, where $//$ denotes the
sequence of rational curves with self-intersection numbers $-1$, $-2$, $-2$, $-3$, $-1$, $-5$, $-1$, $-3$, $-2$,
$-2$, $-1$~\cite{Morrison:2012js,Taylor:2012dr}. The rays $v_i$ of $B_2$ are 
\be
\bsp
v_1 & =  (-1, -12)\\
v_2 & = (0, 1)\\
v_3 & = (1, 6)\\
 & \vdots\\
v_{99} & = (0, -1) \,.
\end{split}
\ee
The rays $v_4,\dots,v_{98}$ can be determined by the condition
\be
v_{i -1} + v_{i + 1}+ (C_i \cdot C_i) v_i = 0,
\ee 
where $C_i \cdot C_i$
is the self-intersection of the curve $C_i$ corresponding to the 2D ray $v_i$.
The toric rays of $B_{\rm max}$ are then given by
\be
\bsp
w_0 & =  (0, 0, 1)\\
w_i & =  (v_i, 0)\ ,\ 1 \leq i \leq 99\\
w_{100} & = (84, 492, -1) = (12v_{15}, -1),
\end{split}
\ee
where $C_{15}$ is the single curve in $B_2$ of self-intersection $-11$.  The
3D cones of the fan for $B_{\rm max}$ are $(w_0, w_i, w_{i + 1})$
and $(w_{100}, w_i, w_{i + 1})$, including the cyclic case
$(w_0/w_{100}, w_{99}, w_1)$. The topology of the divisors $D_1,\dots,D_{99}$ are all that of Hirzebruch surfaces $\mb{F}_n$ with various $n$, while the topology of $D_0$ and $D_{100}$ are the same as $B_2$. Especially, the divisor $D_{15}$ has the topology of $\mb{F}_0=\mb{P}^1\times\mb{P}^1$. The toric divisor $D_{15}$ of $B_{\rm max}$ has an $E_8$ gauge group with a (4,6) curve. Repeating the analysis in section~\ref{sec:btp}, we find that there is a single $E_8\times\varnothing$ type (4,6) curve with the topology of $\mb{P}^1$. 

\subsection{6D origin of the world volume theory on (4,6) curves}
\label{sec:6Dorigin}

We are going to present a rough classification of the 4D theory $\mc{T}_\Sigma$ localized on a (4,6) curve $\Sigma$ based on their local Weierstrass form. For the complete intersection curve $\Sigma=D_1\bigcap D_2$, we denote the local hypersurface equations of $D_1$ and $D_2$ by $u=0$ and $v=0$ in the local coordinate patch $(u,v,w)$ on $B_3$. The gauge groups on $D_1$ and $D_2$ are denoted as $G_1$ and $G_2$\footnote{Here the notation $G_2$ denotes an arbitrary Lie group, not the exceptional Lie group $G_2$}.

For a polynomial in variables $(x_1,\dots,x_n)$:
\be
f=\sum_{m\in S}f_m\prod_{i=1}^n x_i^{a_{m,i}},\ (f_m\neq 0),
\ee
where $S$ is the set of monomials in $f$, we define the notion \textit{lowest order terms in $(x_1,\dots,x_k)$} to be the subset $S_L\subset S$ of monomials:
\begin{equation}
	\begin{split}
	S_L&=\{m=\prod_{i=1}^n x_i^{a_{m,i}}\in S|\forall m'=\prod_{i=1}^n x_i^{a_{m',i}}\in S/\{m\}\ ,\ \\
	&\exists m\leq k, \mathrm{s.\ t.\ }a_{m',i}>a_{m,i}\}
	\end{split}
\end{equation}

Then we separate the (4,6) curves into two classes depending on the functional form of the discriminant locus $\Delta$.

\vspace{0.1in}
\noindent
 
(1) \textit{Class I (4,6) curve}: the the lowest order terms in $(u,v)$ of the discriminant polynomial $\Delta=4f^3+27g^2$ does not depend on $w$.

For example, the following $E_6\times E_6$ model is a class I (4,6) curve:
\be
y^2=x^3+f_3(w)u^3 v^3 x+u^4 v^4,\label{classI}
\ee
as the lowest order term in the discriminant polynomial $\Delta(u,v)$ is $27u^8 v^8$, which does not depend on $w$.

\vspace*{0.1in}
 \noindent

(2) \textit{Class II (4,6) curve}: the lowest order terms in $(u,v)$ of the discriminant polynomial $\Delta=4f^3+27g^2$ depends on $w$.

For example, the following $F_4\times F_4$ model is a class II (4,6) curve:
\be
y^2=x^3+f_3(w)u^3 v^3 x+(1+w)u^4 v^4,\label{classII}
\ee
as the lowest order term in the discriminant polynomial in $(u,v)$ variables is $27u^8 v^8 (1+w)^2$.

\vspace*{0.1in}
 \noindent

\begin{figure}
\centering
\includegraphics[width=8cm]{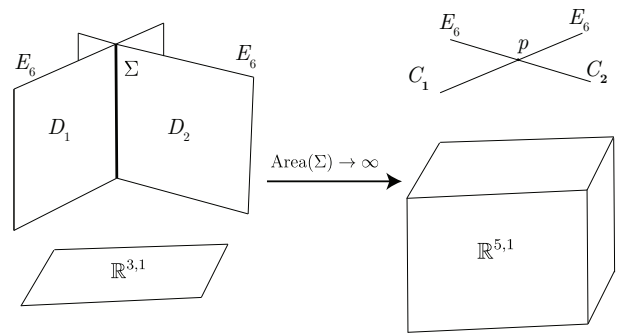}
\caption[E]{\footnotesize In the left picture, we have a 4D F-theory geometry with two stack of $E_6$ branes intersecting each other at the complex curve $\Sigma$. In the decoupling limit, the localized 4D theory $\mc{T}_\Sigma$ becomes a 6D theory in $\mb{R}^{5,1}$, see the right picture. This 6D theory is exactly the 6D $(E_6,E_6)$ SCFT localized at the intersection point $p$ of two curves carrying $E_6$ gauge groups.}\label{f:worldvolume}
\end{figure}

To study the 4D localized theory $\mc{T}_\Sigma$, let us consider the decoupling limit
\be
\mathrm{Vol}(D_1)\ ,\ \mathrm{Vol}(D_2)\ ,\ \mathrm{Area}(\Sigma)\rightarrow\infty.
\ee
Since the divisors $D_1$ and $D_2$ are non-compact, the gauge symmetries $G_1$ and $G_2$ become flavor symmetries. As $\Sigma$ becomes non-compact, the 4D theory $\mc{T}_\Sigma$ becomes a 6D theory in $\mb{R}^{5,1}$, see figure~\ref{f:worldvolume} for a graphic representation.

Then for class I (4,6) curves, such as (\ref{classI}), the 6D theory in the decoupling limit has $E_6\times E_6$ global symmetry, which is precisely the 6D $(E_6,E_6)$ SCFT given by the Weierstrass model\cite{Heckman:2013pva,DelZotto:2014hpa,Heckman:2015bfa}:
\be
y^2=x^3+f_3 u^3 v^3 x+u^4 v^4.\label{E6E6}
\ee
Note that the $f$ polynomial in the above Weierstrass model is irrelevant, as it does not contribute to the lowest order terms in $\Delta$. Here $u=0$ and $v=0$ in (\ref{E6E6}) corresponds to the curves $C_1$ and $C_2$ in figure~\ref{f:worldvolume}.

On the other hand, for class II (4,6) curves, such as (\ref{classII}), the 6D theory has a non-trivial axiodilaton profile on the complex curve $\Sigma$, which breaks the 6D translational symmetry\footnote{In some F-theory models, the $j$-function $j=4f^3/\Delta$ is not well-defined on the curve $u=v=0$~\cite{Esole:2011sm,Halverson:2016vwx}. Nonetheless, the similar statement holds if the lowest order terms of $\Delta$ depnds on $w$.}. In this sense, the matter spectrum for a class II (4,6) curve does not have a simple 6D SCFT origin.

For more general $G_1$ and $G_2$, the 4D theory $\mc{T}_\Sigma$ on a class I (4,6) curve $\Sigma$ is exactly the compactification of a 6D $(G_1,G_2)$ SCFT on $\Sigma$. After $D_1$, $D_2$ and $\Sigma$ retain finite size, $\mc{T}_\Sigma$ couples to the gauge symmetry $G_1$ and $G_2$ again.

For the cases in section~\ref{sec:btp} and ~\ref{sec:max}, such as $(G_1,G_2)=(E_8,\varnothing)$, the 4D theory $\mc{T}_\Sigma$ is a rank-1 E-string theory compactified on $\Sigma$ if $\Sigma$ is a class I (4,6) curve. This is indeed the case for the (4,6) curve on the last $E_8$ in table~\ref{t:cod2} and also the (4,6) curve on $B_{\rm max}$. The local Weierstrass model  has the form of
\be
y^2=x^3+f_4 u^4x+g_5 u^5 v
\ee
near the curve $\Sigma:u=v=0$. Note that the genus $g_\Sigma=0$ in both of these cases.

In the latter parts of this paper, we are mostly focusing on this case where the 4D charged spectrum comes from a 6D E-string theory compactified on $\mb{P}^1$. Of course, the cases of other gauge groups $(G_1,G_2)$ and the cases of class II (4,6) curve are also extremely interesting to study, but we are going to leave this to future work.

In order to compute the 4D particle spectrum localized on the (4,6) curve $\Sigma$, we need to study the spectrum of the E-string theory and its reduction on $\mb{P}^1$. 

\section{Particle spectrum from 6D E-string theory}
\label{sec:spectrum}

6D E-string theory is a strongly coupled (1,0) SCFT whose fundamental object is the tensionless string. One should then expect an infinite tower of massless fields as the excitations of the tensionless string~\cite{Morrison:1996pp}, although there is no notion of asymptotic particle state in the theory~\cite{Chang:2018xmx}. There have been early approaches to study the particle spectrum of E-string theory using an M-theory description, where M2 brane wrapping modes over 2-cycles lead to the electrically charged states under $E_8$ flavor symmetry~\cite{Klemm:1996hh,Morrison:1996pp}. We will employ this philosophy and examine the non-flat fibration of the resolved elliptic Calabi-Yau threefold $\hat{X}_3$ in full details. Finally, we will be able to construct a massless higher spin tower in various representations of $E_8$.

\subsection{Resolution of a non-flat Weierstrass model}
\label{sec:resolution}

In this section, we only consider the singular Weierstrass model $X_3$ over a 2D toric base $B_2$:
\be
y^2=x^3+fxz^4+gz^6,\label{WeierstrassP}
\ee
where $(x:y:z)$ are homogeneous coordinates of the weighted projective space $\mb{P}^{2,3,1}$.

Our general analysis of the resolution of a singular Weierstrass model $X_3$ involves the following steps:

\begin{enumerate}
\item{Constructing a 4D toric ambient space $T$ which is a $\mb{P}^{2,3,1}$ bundle over $B_2$. Denote the toric rays of $B_2$ by $v_1$, $v_2$, $\dots$, $v_n$, and the set of 2D cones in $B_2$ by $S_2$. For the case of a generic elliptic fibration $X_3$ over $B_2$, the 4D rays of $T$ are simply given by:

\be
\bsp
&\tilde{v}_i=(v_i,-2,-3)\ ,\ (i=1,\dots,n)\\
&\tilde{v}_{n+1}=(0,0,1,0)\ ,\ \tilde{v}_{n+2}=(0,0,0,1)\ ,\\
&\tilde{v}_{n+3}=(0,0,-2,-3)
\end{split}
\ee

The set of 4D cones is given by:
\be
\bsp
S_4=\bigcup_{v_i v_j\in S_2}\{&\tilde{v}_i \tilde{v}_j \tilde{v}_{n+1} \tilde{v}_{n+2}\ ,\ \tilde{v}_i \tilde{v}_j \tilde{v}_{n+1} \tilde{v}_{n+3}\ ,\\
	&\tilde{v}_i \tilde{v}_j \tilde{v}_{n+2} \tilde{v}_{n+3}\}.
\end{split}
\ee

The Weierstrass model $X_3$ is then an anti-canonical hypersurface of $T$:
\be
X_3=-K_T.
\ee 
Note that $\tilde{v}_{n+1}$, $\tilde{v}_{n+2}$ and $\tilde{v}_{n+3}$ corresponds to the hypersurface equations $x=0$, $y=0$ and $z=0$ in (\ref{WeierstrassP}) respectively. 

}

\item{Resolving the singular Weierstrass model by blowing up toric subvarieties of the ambient space $T$. We denote a blow up of the intersection locus of $k$ divisors $x_1=x_2=\dots=x_k=0$ by $(x_1,x_2,\dots,x_k;\xi)$, where $\xi=0$ is the exceptional divisor $\hat{E}$ of this blow up~\cite{Lawrie:2012gg}. The Weierstrass form (\ref{WeierstrassP}) is transformed by replacing $x_i\rightarrow x_i\xi$, and then remove a common factor $\xi^m$ from the equation. The condition of crepant resolution is then equivalent to
\be
k=m+1.
\ee

The reason is that the anti-canonical line bundle of the 4D toric ambient space $T'$ is
\be
-K_{T'}=-K_T-(k-1)\hat{E}
\ee
after the blow up. On the other hand, the equation (\ref{WeierstrassP}) is transformed to a section of $-K_T-m\hat{E}$ after the common factor $\xi^m$ is removed. The Calabi-Yau condition then requires that the transformed equation is also a section of the new anti-canonical bundle $-K_{T'}$. Hence we can derive $k=m+1$ from the equation
\be
-K_{T'}=-K_T-m\hat{E}.
\ee

For example, the resolution $(x,y,u;\xi)$ of the following singular Weierstrass model
\be
y^2=x^3+ux+u^2
\ee
is crepant.

}

\item{After the resolution process is done, we get the resolved Calabi-Yau threefold $\hat{X}_3$ as the anti-canonical hypersurface of the 4D ambient space $T_{\rm res}$. We can study the geometry of $\hat{X}_3$ by computing the triple intersection numbers of divisors. Denote the toric divisors in the 4D ambient toric variety $T_{\rm res}$ by $\hat{D}_i$, and the corresponding divisor in the Calabi-Yau hypersurface $\hat{X}$ by $D_i$, then the triple intersection number of $D_i$ are:
\be
D_i\cdot D_j\cdot D_k=-K(T_{\rm res})\cdot\hat{D}_i\cdot\hat{D}_j\cdot\hat{D}_k.\label{tripint}
\ee
Here
\be
-K(T_{\rm res})=\sum \hat{D}_i
\ee
is the anticanonical divisor of $T_{\rm res}$, and the intersection numbers of four toric divisors on $T_{\rm res}$ can be computed by linear equivalence relations, see Appendix~\ref{app:selfint}.

}

\end{enumerate}

To study the higher spin tower of the E-string theory, we choose a generic elliptic fibration over the Hirzebruch surface $\mb{F}_{11}$. $\mb{F}_{11}$ has a toric cyclic representation $(0,-11,0,11)$ and the following 2D toric rays:
\be
v_1=(1,0)\ ,\ v_2=(0,-1)\ ,\ v_3=(-1,-11)\ ,\ v_4=(0,1).
\ee
We can then construct the 4D toric ambient space with rays
\be
\bsp
&\tilde{v}_1=(1,0,-2,-3)\ ,\ \tilde{v}_2=(0,-1,-2,-3)\ ,\\
&\tilde{v}_3=(-1,-11,-2,-3)\ ,\ \tilde{v}_4=(0,1,-2,-3)\ ,\ \\
&\tilde{v}_5=(0,0,1,0)\ ,\ \tilde{v}_6=(0,0,0,1)\ ,\ \tilde{v}_7=(0,0,-2,-3)
\end{split}
\ee
as discussed earlier. 

Denote the local equations of $v_1$ and $v_2$ on the base $\mb{F}_{11}$ by $v=0$ and $u=0$ respectively, the singular Weierstrass model $X_3$ can be written as:
\be
\bsp
y^2=&x^3+f_4(v) u^4 x^4+(g_{5,0}+g_{5,1}v)u^5 z^6\\
&+(\mathrm{higher\ order\ terms}).
\end{split}
\ee
One can see that there is a (4,6) point at $u=g_{5,0}+g_{5,1}v=0$.

The resolution process of an $E_8$ Weierstrass model 
\be
y^2=x^3+b_4 u^4 x+b_6 u^5\label{E8w}
\ee
was studied in~\cite{Lawrie:2012gg}. We list the sequence of blow ups and the 4D rays corresponding to each of the exceptional divisors in table~\ref{t:blpE8}. For completeness, we also list all the 4D rays and cones of the 4D toric ambient space $T_{\rm res}$ in appendix~\ref{app:toricfan}. After this sequence of blow ups, the Weierstrass model (\ref{E8w}) is transformed into
\be
\bsp
\delta_1\delta_2\epsilon_4\epsilon_7 y^2&=\zeta_1\zeta_2\epsilon_1\epsilon_5(\delta_2\zeta_2\epsilon_3^2\epsilon_8 x^3+u^4\zeta_1^2\delta_1^2\epsilon_1\epsilon_2^4\epsilon_4\epsilon_6^2\epsilon_9\\
&(b_4x\zeta_2\delta_2\epsilon_1\epsilon_3^2\epsilon_4\epsilon_5^2\epsilon_6^2\epsilon_7^2\epsilon_8^3\epsilon_9^3\epsilon_{10}^4+b_6 u)).\label{E8resolved}
\end{split}
\ee

\begin{table}
\centering
\begin{tabular}{|c|c|}
\hline
The blow up&The 4D toric ray\\
\hline
$(x,y,u;\zeta_1)$ & $(0,-1,-1,-2)$\\
$(x,y,\zeta_1;\zeta_2)$ & $(0,-1,0,-1)$\\
$(y,\zeta_1;\delta_1)$ & $(0,-1,-1,-1)$\\
$(y,\zeta_2;\delta_2)$ & $(0,-1,0,0)$\\
$(\zeta_2,\delta_1;\epsilon_1)$ & $(0,-2,-1,-2)$\\
$(\zeta_1,\delta_1;\epsilon_2)$ & $(0,-2,-2,-3)$\\
$(\zeta_2,\delta_2;\epsilon_3)$ & $(0,-2,0,-1)$\\
$(\delta_1,\delta_2;\epsilon_4)$ & $(0,-2,-1,-1)$\\
$(\delta_2,\epsilon_1;\epsilon_5)$ & $(0,-3,-1,-2)$\\
$(\epsilon_1,\epsilon_4;\epsilon_6)$ & $(0,-4,-2,-3)$\\
$(\delta_2,\epsilon_4;\epsilon_7)$ & $(0,-3,-1,-1)$\\
$(\delta_2,\epsilon_5;\epsilon_8)$ & $(0,-4,-1,-2)$\\
$(\epsilon_4,\epsilon_5;\epsilon_9)$ & $(0,-5,-2,-3)$\\
$(\epsilon_5,\epsilon_7;\epsilon_{10})$ & $(0,-6,-2,-3)$\\
\hline
\end{tabular}
\caption{The full blow up sequence that resolves an $E_8$ singular Weierstrass model and the correspondence with the 4D rays in the toric ambient space $T_{\rm res}$.}\label{t:blpE8}
\end{table}

The following irreducible exceptional divisors correspond to the eight Dynkin nodes of $E_8$:
\be
\bsp
\{&\epsilon_2=0\ ,\ \delta_1=0\ ,\ \epsilon_6=0\ ,\ \epsilon_9=0\ ,\\
&\epsilon_{10}=0\ ,\ \epsilon_8=0\ ,\ \epsilon_3=0\ ,\ \epsilon_{7}=0\}.
\end{split}
\ee

We rename them to $E_1,E_2,\dots,E_8$, and their intersection relations can be computed by inspecting the topology of the vertical divisor $v=0$, which is labeled as $D_1$ with 4D toric ray $(1,0,-2,-3)$. The non-vanishing triple intersection numbers involving $D_1$ are
\be
\bsp
&D_1\cdot D_5\cdot D_6=7\ ,\ D_1\cdot D_4\cdot D_5=2\ ,\ D_1\cdot D_5\cdot E_7=1\ ,\\
&D_1\cdot D_4\cdot D_6=3\ ,\ D_1\cdot D_6\cdot E_8=1\ ,\ D_1\cdot D_4\cdot D_7=1\ ,\\
&D_1\cdot D_2\cdot D_7=1\ ,\ D_1\cdot D_2\cdot E_1=1\ ,\ D_1\cdot E_1\cdot E_2=1\ ,\\
&D_1\cdot E_2\cdot E_3=1\ ,\ D_1\cdot E_3\cdot E_4=1\ ,\ D_1\cdot E_4\cdot E_5=1\ ,\\
&D_1\cdot E_5\cdot E_6=1\ ,\ D_1\cdot E_5\cdot E_8=1\ ,\ D_1\cdot E_7\cdot E_8=1\ ,\\
&D_1 D_5^2=4\ ,\ D_1 D_6^2 =10\ ,\\
&D_1 D_7^2=D_1 D_2^2=D_1 E_i^2=-2\ (i=1,\dots,8).
\end{split}
\ee

\begin{figure}
\centering
\includegraphics[height=2cm]{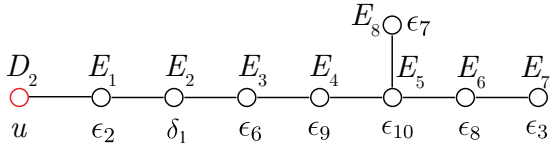}
\caption[E]{\footnotesize The intersection relation of curves on the vertical divisor $D_1$. Each node denotes a curve which is the intersection of $D_1$ with the corresponding divisor $D$. These curves then form an affine $E_8$ Dynkin diagramon $D_1$. The exceptional divisors are labelled by $E_1,E_2,\dots,E_8$.}\label{f:E8resol}
\end{figure}

From the intersection number of the vertical divisor $D_1$ with the exceptional divisors $E_i$, we see that the exceptional divisors form an affine Dynkin diagram of $E_8$ in figure~\ref{f:E8resol}. Since we have $D_1 E_i^2=-2$, $D_1^2 E_i=0$, we can easily check that the exceptional curve $D_1\cdot E_i$ is indeed a rational curve by the adjunction formula on a Calabi-Yau threefold and Riemann-Roch theorem:
\be
2g(D_1\cdot E_i)-2=D_1 E_i^2+D_1^2 E_i=-2.
\ee

Besides these exceptional divisors $E_1,\dots,E_8$, there is also a non-flat fiber component on $\hat{X}_3$. It corresponds to the equation $\delta_2=0$ and the 2D toric ray $(0,-1,0,0)$, and we denote it by $S_{nf}$. As one can see in equation (\ref{E8resolved}), the equation 
\be
\delta_2=b_6=0
\ee
describes a complex surface localized over the (4,6) point on the base.

The non-vanishing triple intersection numbers involving $S_{nf}$ are
\be
\bsp
&S_{nf}\cdot D_5\cdot D_6=1\ ,\ S_{nf}\cdot D_5\cdot E_7=1\ ,\ S_{nf}\cdot D_6\cdot E_8=1\ ,\\
&S_{nf}\cdot E_7\cdot E_6=1\ ,\ S_{nf}\cdot E_8\cdot E_6=1\ ,\ S_{nf}^2 D_5=-2\ ,\\
&S_{nf}^2 D_6=-3\ ,\ S_{nf}^2 E_6=S_{nf}^2 E_8=-1\ ,\ S_{nf} D_6^2=1\ ,\\
&S_{nf} E_7^2=-2\ ,\ S_{nf} E_8^2=-1\ ,\ S_{nf} E_6^2=-1\ ,\ S_{nf}^3=7.
\end{split}
\ee

\begin{figure}
\centering
\includegraphics[height=4cm]{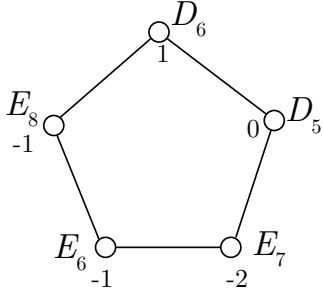}
\caption[E]{\footnotesize The intersection of curves on the non-flat fiber $S_{nf}$. Each node denotes a rational curve which is the intersection of $S_{nf}$ with the corresponding divisor $D$. The number on a node corresponds to the triple intersection number $S_{nf}\cdot D^2$, and the edges between nodes correspond to an intersection point.}\label{f:Snf}
\end{figure}

We plot the intersection relations of the curves on $S_{nf}$ in Figure~\ref{f:Snf}. Using the adjunction formula on $\hat{X}_3$, we can indeed check that the curves $S_{nf}\cdot D_5$, $S_{nf}\cdot D_6$, $S_{nf}\cdot E_6$, $S_{nf}\cdot E_7$ and $S_{nf}\cdot E_8$ are all rational. We can hence see that the non-flat fiber $D(\delta_2)$ has the topology of a generalized del Pezzo surface $gdP_2$ with toric cyclic representation $(1,-1,-1,-2,0)$. As a crosscheck, note that for any rational surface $D$ in an arbitrary complex threefold $X$, we have the relation
\be
\bsp
h^{1,1}(D)&=10-K_D^2\\
&=10-D\cdot (K_X+D)^2.
\end{split}
\ee
For the case of a Calabi-Yau threefold $X$ with $K_X=0$, we have
\be
h^{1,1}(D)=10-D^3.\label{h11DCY}
\ee
Since the self-triple intersection number of $S_{nf}$ is $S_{nf}^3=7$, we have indeed confirmed that $S_{nf}$ is a rational surface with $h^{1,1}(S_{nf})=3$\footnote{Since $S_{nf}$ is fully contractable to a point, this surface has to be rational~\cite{Morrison:1996pp}}. Along with the intersection structure of rational curves on $S_{nf}$, we have proved that $S_{nf}$ indeed has the topology of a $gdP_2$.

\begin{figure}
\centering
\includegraphics[height=3cm]{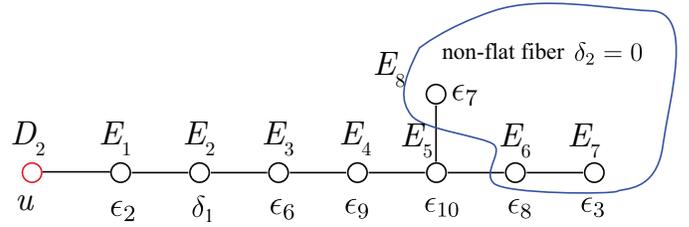}
\caption[E]{\footnotesize The intersection of the non-flat fiber with the exceptional divisors of $E_8$.}\label{f:E8nf}
\end{figure}

We also plot the intersection structure of the non-flat fiber with the exceptional divisors of $E_8$ in Figure \ref{f:E8nf}. 

\begin{figure}
\centering
\includegraphics[height=6cm]{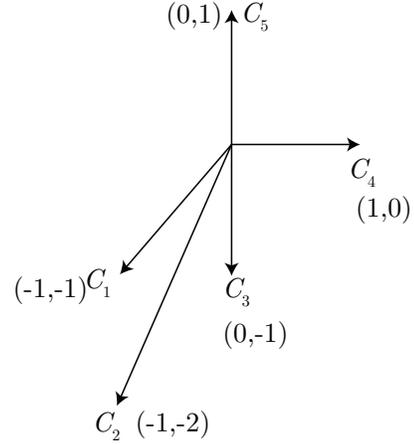}
\caption[E]{\footnotesize Toric diagram of generalized del Pezzo surface $gdP_2$.}\label{f:gdP2}
\end{figure}

We plot the toric diagram of $gdP_2$ in Figure~\ref{f:gdP2}, where we have identified $C_1=S_{nf}\cdot E_8$, $C_2=S_{nf}\cdot E_6$, $C_3=S_{nf}\cdot E_7$, $C_4=S_{nf}\cdot D_5$ and $C_5=S_{nf}\cdot D_6$. The three negative rational curves $C_1$, $C_2$ and $C_3$ generates the Mori cone of $gdP_2$. 

Since the divisor $S_{nf}$ does not intersect with the horizontal divisor $D_7$ or vertical divisors $D_1$, $D_3$ at all, one can again see that $S_{nf}$ is entirely in the fiber direction. Before we continue to the discussion of M2 brane spectrum in the M-theory picture, we proceed to check the topology of other important divisors $D\subset \hat{X}_3$ using the same methodology.

For the horizontal divisor $D_7$, the non-vanishing triple intersection numbers are
\be
\bsp
&D_7\cdot D_1\cdot D_2=D_7\cdot D_2\cdot D_3=D_7\cdot D_3\cdot D_4 \\
&\qquad\qquad\qquad\qquad\qquad\qquad =D_7\cdot D_4\cdot D_1=1\\
&D_7\cdot D_1^2=D_7\cdot D_3^2=0\ ,\ D_7\cdot D_2^2=-11\ ,\ D_7\cdot D_4^2=11\\
&D_7^2\cdot D_1=D_7^2\cdot D_3=-2\ ,\ D_7^2\cdot D_2^2=9\ ,\ D_7^2\cdot D_4=-13\ ,\\
&D_7^3=8.
\end{split}
\ee
From these numbers, one can see that $D_7$ indeed has the topology of the Hirzebruch surface $\mb{F}_{11}$ with $h^{1,1}(D_7)=2$. Hence our resolved Calabi-Yau threefold $\hat{X}_3$ is indeed an elliptic fibration over the original base $\mb{F}_{11}$.

For the exceptional divisors $E_1,\dots,E_8$, it turns out that $E_1$, $E_2$, $E_3$, $E_4$, $E_5$ and $E_7$ are all Hirzebruch surfaces with topology of $\mb{F}_7$, $\mb{F}_5$, $\mb{F}_3$, $\mb{F}_1$, $\mb{F}_1$ and $\mb{F}_4$ respectively. The 0-curve on these Hirzebruch surfaces are precisely the intersection of $E_i$ with the vertical divisors $D_1$ and $D_3$, which are the exceptional $\mb{P}^1$ curves giving rise to gauge bosons.

\begin{figure}
\centering
\includegraphics[width=0.45\textwidth]{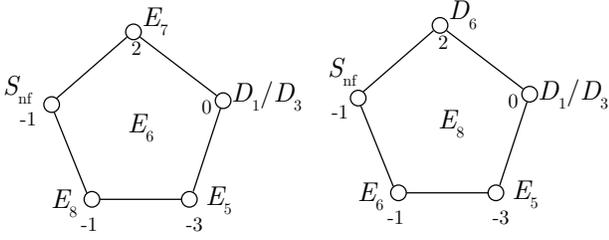}
\caption[E]{\footnotesize The intersection of curves on the exceptional divisors $E_6$ and $E_8$, which both have the topology of a toric surface $(2,-1,-1,-3,0)$ with $h^{1,1}=3$.}\label{f:E6E8}
\end{figure}

The exceptional divisor $E_6$ and $E_8$ all have the topology of a toric surface $(2,-1,-1,-3,0)$ with $h^{1,1}=3$, see Figure~\ref{f:E6E8} for the intersection relations of rational curves. The 0-curve on $E_6$ and $E_8$ represents the exceptional $\mb{P}^1$ curve as well. 

\subsection{E-string spectrum from M2 brane wrapping modes}
\label{sec:6Dspectrum}

Now we are ready to consider the M2 brane wrapping mode on a complex curve $\Sigma \subset S_{nf}$. We will apply the methodology of~\cite{Witten:1996qb,Kachru:2018nck} on our example.

The 8 real supercharges in the 5D $\mc{N}=1$ supergravity are under the representation of $2\cdot (1/2,0)\oplus 2\cdot(0,1/2)$ of the 5D massive little group $SO(4) \simeq SU(2)_L\times SU(2)_R$. After the M2 brane is included, the supercharges $2\cdot (0,1/2)$ are broken, and we get 4 fermionic zero modes after we act them on the ground state:
\be
H_0 = 2(0, 0)\oplus (0, 1/2).
\ee
Besides that, we also need to consider the moduli space $\mathcal{M}_\Sigma$ of the curve $\Sigma$. The four unbroken supercharges $2\cdot(1/2,0)$ exactly correspond to the four differential operators $\partial$, $\bar\partial$, $\partial^*$ and $\bar{\partial}^*$ on the Kahler manifold $\mc{M}_\Sigma$. Hence the BPS states exactly correspond to the harmonic forms on $\mc{M}_\Sigma$. We then naturally identify $SU(2)_L$ as the Lefschetz decomposition SU(2) acting on the cohomology groups of $\mc{M}_\Sigma$. Especially, the $SU(2)_L$ generator $J_3$ acts on a differential form $|\psi\rangle$ as
\be
J_3|\psi\rangle=\frac{(p+q-\text{dim}_\mathbb{C}\mathcal{M}_\Sigma)}{2}|\psi\rangle\ ,\ |\psi\rangle\in H^{p,q}(\mc{M}_\Sigma).
\ee
Hence the classes of the harmonic forms $\omega_{p,q}$ and $\omega_{N-p,N-q}$ on $\mathcal{M}_\Sigma$ together give rise to a particle state in the representation $(|n|/2, 0)$ under the little group $SO(4)$, where $n = p + q -\text{dim}_\mathbb{C}\mathcal{M}_\Sigma $, $N = \text{dim}_\mathbb{C}\mathcal{M}_\Sigma$ and $p + q \neq N$. When $p + q = N$, the harmonic form $\omega_{p,q}$ gives rise to a real scalar field in the representation $(0,0)$. In order to avoid double counting the state, we should always restrict to the upper half the Hodge diamond of $\mathcal{M}_\Sigma$, i. e., $p + q \leq N$. Tensor together with the half-hypermultiplet $H_0$, the resulting  particle spectrum contains the states in the representation:
\be
\bsp
	&[2(0,0)\oplus \left(0,\frac{1}{2}\right)]\otimes[\sum\limits_{p + q \leq N}h^{p,q}\cdot\left(\frac{|n|}{2},0\right)]\\
	&=\sum\limits_{p + q \leq N} 2h^{p,q}\cdot\left(\frac{|n|}{2}, 0\right)\oplus h^{p,q}\cdot\left(\frac{|n|}{2}, \frac{1}{2}\right).
\end{split}
\ee

Now we need to compute the Hodge numbers $h^{p,q}$ of $\mathcal{M}_\Sigma$. We consider a complex curve $\Sigma$ in the form of a complete intersection of two divisors $D_a$ and $D_b$ of the Calabi-Yau threefold: $\Sigma=D_a\cdot D_b$. We define $\Sigma_a\subset D_a$ and $\Sigma_b\subset D_b$ to be the effective divisors corresponding to $\Sigma$ on $D_a$ and $D_b$ respectively. Then the moduli space $\mathcal{M}_\Sigma$ of the curve $\Sigma$ is a product of the linear system $|\Sigma_a|$ on $\Sigma_a$ and $|\Sigma_b|$ on $\Sigma_b$:
\be
\bsp
	\mathcal{M}_\Sigma &= \mb{P}^{\dim(|\Sigma_a|)}\times\mb{P}^{\dim(|\Sigma_b|)}\\
	&=\mb{P}^{h^0(\Sigma_a,D_a) - 1}\times\mb{P}^{h^0(\Sigma_b,D_b) - 1}.
\end{split}
\ee
The dimension of the linear system $\Sigma_a$ on $D_a$ can be computed by
\be
\bsp
\dim(|\Sigma_a|)&=h^0(\Sigma_a,D_a)-1\\
&=\max\left(\frac{\Sigma_a\cdot \Sigma_a-K_a\cdot \Sigma_a}{2},0\right)\\
&=\max(\Sigma_a\cdot \Sigma_a-g_\Sigma,0),
\end{split}
\ee
where $K_a$ is the canonical divisor of $D_a$ and we have used the Riemann-Roch formula on $D_a$. Similarly, we have
\be
\bsp
\dim(|\Sigma_b|)&=h^0(\Sigma_b,D_b)-1\\
&=\max\left(\frac{\Sigma_b\cdot \Sigma_b-K_b\cdot \Sigma_b}{2},0\right)\\
&=\max(\Sigma_b\cdot \Sigma_b-g_\Sigma,0).
\end{split}
\ee
Using the adjunction formula on the Calabi-Yau threefold $\hat{X}_3$, we have:
\be
\bsp
\Sigma_a\cdot \Sigma_a+\Sigma_b\cdot \Sigma_b&=D_a\cdot D_b\cdot (D_a+D_b)\\
&=2g_\Sigma-2.
\end{split}
\ee
Hence we can see that it is impossible to have $\dim(|\Sigma_a|)>0$ and $\dim(|\Sigma_b|)>0$ at the same time. In other words, the moduli space $\mc{M}_\Sigma$ always only has one component $\mathbb{P}^N$ with Hodge numbers
\be
	h^{p,q}=\delta_{p,q}\  (0\leq p,q\leq N)
\ee
Therefore, for fixed $N=\dim_{\mb{C}}(\mc{M}_\Sigma)$, all the states together give rise to the following 5D massive supermultiplet:
\be
	R_N=\left(\frac{N}{2}, \frac{1}{2}\right)\oplus2\left(\frac{N}{2}, 0\right).\label{Rn}
\ee
After going to the F-theory limit, the above massive supermultiplet becomes a massless supermultiplet in 6D transforming under the same little group $SO(4)\simeq SU(2)_L\times SU(2)_R$ of the 6D massless particles.

Apart from the representation of Lorentz group, we also need to compute the representation of an M2 brane wrapping mode under the $E_8$ gauge group. For a single M2 brane wrapping $\Sigma$, its electric charge under each Cartan subgroup $U(1)_i\subset E_8$ is given by the intersection number:
\be
q_i=\Sigma\cdot E_i.
\ee
Hence this M2 brane wrapping mode is associated to an $E_8$ weight vector
\be
\vec{w}(\Sigma)=(q_1,q_2,\dots,q_8)
\ee
in a representation $R$ of $E_8$. The other states in the representation $R$ are given by M2 brane wrapping modes on the union of $\Sigma$ and other exceptional $\mb{P}^1$s (possibly wraps an extra $\mathbb{P}^1$ multiple times.). The labeling of Dynkin nodes can be found in figure~\ref{f:E8resol} or figure~\ref{f:E8}. We also list a number of facts about the $E_8$ Lie algebra in appendix~\ref{app:group}.

Now we are going to list the curve classes on $S_{nf}$ with the lowest dimensional moduli spaces $\mathrm{dim}_{\mb{C}}(\mc{M}_\Sigma)$, which corresponds to 6D massless multiplets with the lowest spins. We denote the local equation of the toric curve $C_i\subset S_{nf}$ by $u_i=0$. 

\begin{enumerate}
\item{$R_0=(0,1/2)+2(0,0)$, $\mathrm{dim}_{\mb{C}}(\mc{M}_\Sigma)=0$: $\Sigma=C_1$, $C_2$

The only curves on $S_{nf}$ with $\mathrm{dim}_{\mb{C}}(\mc{M}_\Sigma)=0$ are the rational (-1)-curves $C_1$ and $C_2$. Their non-zero intersection numbers with the exceptional divisors are:
\be
\bsp
&C_1\cdot E_8=-1\ ,\ C_1\cdot E_6=1\ ,\ C_2\cdot E_8=1\ ,\\
&C_2\cdot E_6=-1\ ,\ C_2\cdot E_7=1.
\end{split}
\ee
Hence the weight vectors corresponding to $C_1$ and $C_2$ are
\be
\bsp
&\vec{w}(C_1)=(0,0,0,0,0,1,0,-1)\ ,\\
&\vec{w}(C_2)=(0,0,0,0,0,-1,1,1),
\end{split}
\ee
which are both in the adjoint representation $\mbf{248}$ of $E_8$.

This suggests the appearance of a 6D hypermultiplet $R_0$ in the representation $\mbf{248}$, which is consistent with the results in~\cite{Klemm:1996hh}.

}

\item{$R_1=(1/2,1/2)+2(1/2,0)$,  $\mathrm{dim}_{\mb{C}}(\mc{M}_\Sigma)=1$: $\Sigma=C_3$, $C_1+C_2$

The curve classes with $\mathrm{dim}_{\mb{C}}(\mc{M}_\Sigma)=1$ include the rational curves with self-intersection (-2) or 0. The only (-2)-curve on $S_{nf}$ is the curve $C_3$, which is itself an elliptic $\mb{P}^1$. Hence M2 brane wrapping mode on $C_3$ merely corresponds to a W-boson of the $E_8$ gauge group, which is not anything new in the E-string spectrum.

The only other irreducible curve class with $\mathrm{dim}_{\mb{C}}(\mc{M}_\Sigma)=1$ is $\Sigma=C_1+C_2$. The equation of a generic curve in the divisor class $C_1+C_2$ is
\be
u_1 u_2+a u_4=0,
\ee
where $a$ is a non-zero complex number. Note that $C_1+C_2$ is equivalent to $C_4$ in Figure~\ref{f:gdP2}. The only non-zero intersection between the curve $C_4$ and the exceptional divisors in Figure~\ref{f:E8nf} is
\be
C_4\cdot E_7=1,
\ee
hence the state from M2 brane wrapping $C_4$ gives rise to the highest weight state $(0,0,0,0,0,0,1,0)$ of the representation $\mbf{3875}$ of $E_8$. Along with other states from M2 brane wrapping unions of $C_4$ and the exceptional $\mathbb{P}^1$s, we will get a 6D vector multiplet $R_1$ in the representation $\mbf{3875}$ of $E_8$.

}

\item{$R_2=(1,1/2)+2(1,0)$ Rarita-Schwinger multiplet $\mathrm{dim}_{\mb{C}}(\mc{M}_\Sigma)=2$: $\Sigma=C_1+2C_2+C_3$

The only irreducible curve class with $\mathrm{dim}_{\mb{C}}(\mc{M}_\Sigma)=2$ is $C_1+2C_2+C_3$, which is equivalent to the curve $C_5$ in figure~\ref{f:gdP2}. The equation of a generic curve in $C_1+2C_2+C_3$ is
\be
u_1 u_2^2 u_3+a u_2 u_3 u_4+b u_5=0
\ee

The only non-zero intersection between the curve $C_5$ and the exceptional divisors  in figure~\ref{f:E8nf} is
\be
C_4\cdot E_8=1.
\ee
The M2 brane wrapping mode on $C_5$ is the highest weight state $(0,0,0,0,0,0,0,1)$ of the representation $\mbf{147250}$. We hence have a 6D Rarita-Schwinger multiplet $R_2$ in the representation $\mbf{147250}$ of $E_8$.

}

\item{$R_3=(3/2,1/2)+2(3/2,0)$,  $\mathrm{dim}_{\mb{C}}(\mc{M}_\Sigma)=3$: $\Sigma=2C_1+2C_2+C_3$

The only irreducible curve class with $\mathrm{dim}_{\mb{C}}(\mc{M}_\Sigma)=3$ is $2C_1+2C_2+C_3$. The equation of a generic curve in $2C_1+2C_2+C_3$ is
\be
u_1^2 u_2^2 u_3+a u_1 u_2 u_3 u_4+b u_3 u_4^2+c u_1 u_5=0
\ee

The only non-zero intersections between the curve $2C_1+2C_2+C_3$ and the exceptional divisors  in figure~\ref{f:E8nf} is
\be
(2C_1+2C_2+C_3)\cdot E_6=1.
\ee
The M2 brane wrapping mode on $2C_1+2C_2+C_3$ is the highest weight state $(0,0,0,0,0,1,0,0)$ of $E_8$. We then have a 6D multiplet $R_3$ with little group representation $(3/2,1/2)\oplus 2(3/2,0)$ in the representation $\mbf{6696000}$ of $E_8$.

}

\end{enumerate}

Hence we have obtained a massless higher spin tower in the E-string theory, where the larger $E_8$ representation corresponds to field with larger spin. The interpretation of these fields as particle states in the language of superconformal field theory is not clear, since E-string theory is strongly coupled~\cite{Chang:2018xmx}. We will return to this problem in section~\ref{sec:discussion}.

\section{4D spectrum of compactified E-string theory}
\label{sec:4D}

In this section, we consider a 4D F-theory scenario of an $E_8$ gauge group on a divisor $S\subset B_3$ with a class I (4,6) curve $\Sigma\subset S$ of $E_8\times\varnothing$ type. The localized matter on $\Sigma$ in the 4D F-theory is precisely the compactification of the 6D rank-1 E-string theory on a complex curve $\Sigma$.

If we simply take the $\Sigma=T^2$ without the inclusion of any gauge flux, the 4D IR fixed point is the rank-1 $E_8$ Minahan-Nemeschansky theory~\cite{Minahan:1996cj,Minahan:1998vr}. Recently, the BPS spectrum of the rank-1 $E_8$ MN theory was obtained by a different geometric approach, using the worldvolume theory of a D3 brane probe~\cite{Distler:2019eky}. Nonetheless, the spectrum of M2 brane wrapping states on rational curves completely agree with our result in the previous section.

More generally, we can consider the compactification of 6D spectrum on a general Riemann surface $\Sigma$ with genus $g$, in presence of gauge flux supported on $\Sigma$. The $E_8$ gauge symmetry (equivalently the $E_8$ flavor symmetry of E-string theory) will be broken to a subgroup $H\times U(1)^r\subset E_8$. For a 6D field in representation $R(E_8)$, the reduced 4D spectrum are labelled by the representations $R_q$ in the branching rule of $R(E_8):E_8\rightarrow H\times U(1)^r$, where $R$ is a representation of $H$ and
\be
q=(q_1,q_2,\dots,q_r)
\ee
is the charge vector under $U(1)^r$.

The flux breaking of the gauge group in F-theory has been studied in~\cite{Beasley:2008dc,Beasley:2008kw,Donagi:2008ca,Donagi:2008kj}. In our case, we consider a vertical $G_4$ flux in the M-theory picture:
\be
G_4=\sum_{i=1}^r F_i\wedge \omega_i,
\ee
where $F_i$ is a (1,1)-form on $S$ and $\omega_i$ is the (1,1)-form Poincar\'{e} dual to the exceptional divisor $E_i$. The Poincar\'{e} dual of $F_i$ on $S$ corresponds to a line bundle $L_i\in \mathrm{Pic}(S)$, which is refered as the bulk gauge bundle associated to the hypercharge $U(1)_i$. 

We denote the ``bulk'' line bundle corresponds to a representation $R_q$ by
\be
L_q=\sum_{i=1}^r q_i L_i,
\ee
and its restriction to $\Sigma$ by
\be
L_\Sigma(q)=\mc{O}(\sum_{i=1}^r q_i L_i\cdot\Sigma).\label{LSigma}
\ee

When $\Sigma$ is not a torus, in order to preserve $\mathcal{N} = 1$ SUSY in 4D, the 6D theory must be twisted since there is no covariantly constant spinor on a Riemann surface $\Sigma$ with $g\neq 1$. The twist is given by~\cite{Beasley:2008dc}:
\be\label{eq:twist}
	J_{top} = J - \frac{1}{2}R
\ee
where $J$ is the charge of the field under the structure group $U(1)$ of $\Sigma$ and $R$ is the charge under the diagonal $U(1)_R$ of the $SU(2)_R$ R-symmetry. To study the reduction of the 6D multiplets $R_n$ in (\ref{Rn}) to 4D fields under the twisting, we need to study the branching rules of 
\be
\bsp
SO(1,5)\times SU(2)_R &\rightarrow SO(1,3)\times U(1)_J\times U(1)_R\\
&\simeq SU(2)_1\times SU(2)_2\times U(1)_J\times U(1)_R.\label{4Dbranch}
\end{split}
\ee
Then for each field component, we can compute its $J_{top}$ which determines its multiplicity along with the gauge bundle (\ref{LSigma}). For the cases of $J_{top}\leq 0$, we have a field component 
\be
\phi\in H^0_{\bar\partial}(K_\Sigma^{-J_{top}}\otimes L_\Sigma(q),\Sigma).\label{negJtop}
\ee
While for the cases of $J_{top}\geq 0$, we have
\be
\bsp
\bar\phi&\in H^0_{\partial}(\overline{K}_\Sigma^{J_{top}}\otimes \overline{L_\Sigma(q)^*},\Sigma)\\
&\cong\overline{H^0_{\bar\partial}(K_\Sigma^{J_{top}}\otimes L_\Sigma(q)^*,\Sigma)}.\label{posJtop}
\end{split}
\ee
Note that we have used the fact the complex conjugate $\bar{E}$ of a vector bundle is isomorphic to its dual bundle $E^*$. In absence of the gauge bundle $L_\Sigma(q)$, two field components with the same $|J_{top}|$ always share the same multiplicity which can be seen via comparing (\ref{negJtop}) and (\ref{posJtop}). 

Now we are going to analyze the reduction of 6D representations $R_n$ (\ref{Rn}) with small $n$ in detail.

\begin{enumerate}
\item{6D hypermultiplet $2R_0=2(0,1/2)\oplus 4(0,0)$\footnote{Note that the representation $R_0=(0,1/2)\oplus 2(0,0)$ is a 6D half-hypermultiplet.}.

This case was already discussed in~\cite{Beasley:2008dc}. In the Lorentz group representation, the fermionic part is the anti-chiral spinor $\mbf{4}'$, while the bosonic part contains two complex scalars in the trivial representation $\mbf{1}$ under the Lorentz group. The $SU(2)_R$ representations of the fermionic and bosonic parts are $\mbf{1}$ and $\mbf{2}$ respectively. The branching rules under (\ref{4Dbranch}) are
\be
(\mbf{4}',\mbf{1})\rightarrow(\mbf{2},\mbf{1},-\frac{1}{2},0)+(\mbf{1},\mbf{2},\frac{1}{2},0),
\ee
\be
(\mbf{1},\mbf{2})\rightarrow (\mbf{1},\mbf{1},0,1)+(\mbf{1},\mbf{1},0,-1)
\ee
After the twisting, the 4D states under $SO(1,3)_{J_{top}}$ are
\be
(\mbf{2},\mbf{1})_{-1/2}+(\mbf{1},\mbf{2})_{1/2}+(\mbf{1},\mbf{1})_{-1/2}+(\mbf{1},\mbf{1})_{1/2}.
\ee
We then write down the field components in representation $R_q$ of $H$:
\be
\bsp
&(\mbf{2},\mbf{1})_{-1/2}: \psi_\alpha(R_q)\in H^0_{\bar\partial}(K_\Sigma^{1/2}\otimes L_\Sigma(q),\Sigma)\\
&(\mbf{1},\mbf{2})_{1/2}: \bar{\psi}_{\dot\alpha}(R_q)\in \overline{H^0_{\bar\partial}(K_\Sigma^{1/2}\otimes L_\Sigma(q)^*,\Sigma)}\\
&(\mbf{1},\mbf{1})_{-1/2}: \phi(R_q)\in H^0_{\bar\partial}(K_\Sigma^{1/2}\otimes L_\Sigma(q),\Sigma)\\
&(\mbf{1},\mbf{1})_{1/2}: \bar\phi(R_q)\in \overline{H^0_{\bar\partial}(K_\Sigma^{1/2}\otimes L_\Sigma(q)^*,\Sigma)}.\label{6Dhyper}
\end{split}
\ee
We can hence see that there are $h^0(K_\Sigma^{1/2}\otimes L_\Sigma(q),\Sigma)$ copies of chiral multiplets and $h^0(K_\Sigma^{1/2}\otimes L_\Sigma(q)^*,\Sigma)$ copies of anti-chiral multiplets in representation $R_q$. We can also write down the field components in the conjugate representation $\bar{R}_{-q}$ of $H$:
\be
\bsp
&(\mbf{2},\mbf{1})_{-1/2}: \psi^c_\alpha(\bar{R}_{-q})\in H^0_{\bar\partial}(K_\Sigma^{1/2}\otimes L_\Sigma(q)^*,\Sigma)\\
&(\mbf{1},\mbf{2})_{1/2}: \bar{\psi}^c_{\dot\alpha}(\bar{R}_{-q})\in \overline{H^0_{\bar\partial}(K_\Sigma^{1/2}\otimes L_\Sigma(q),\Sigma)}\\
&(\mbf{1},\mbf{1})_{-1/2}: \phi^c(\bar{R}_{-q})\in H^0_{\bar\partial}(K_\Sigma^{1/2}\otimes L_\Sigma(q)^*,\Sigma)\\
&(\mbf{1},\mbf{1})_{1/2}: \bar\phi^c(\bar{R}_{-q})\in \overline{H^0_{\bar\partial}(K_\Sigma^{1/2}\otimes L_\Sigma(q),\Sigma)}.
\label{6Dhyperconj}
\end{split}
\ee
Note that we have used $L_\Sigma(-q)=L_\Sigma(q)^*$. Comparing (\ref{6Dhyper}) with (\ref{6Dhyperconj}), we can see that the fermionic fields form CPT conjugates: $(\psi_\alpha(R_q),\bar{\psi}^c_{\dot\alpha}(\bar{R}_{-q})$, $( \bar{\psi}_{\dot\alpha}(R_q),  \psi^c_\alpha(\bar{R}_{-q}))$. Hence we can also interpret the resulting spectrum as $h^0(K_\Sigma^{1/2}\otimes L_\Sigma(q),\Sigma)$ copies of chiral multiplets in representation $R_q$ and $h^0(K_\Sigma^{1/2}\otimes L_\Sigma(q)^*,\Sigma)$ copies of chiral multiplets in representation $\bar{R}_{-q}$. In either way, the net chiral generations of representation $R_q$ is
\be
\bsp
\chi(R_q)&=h^0(K_\Sigma^{1/2}\otimes L_\Sigma(q),\Sigma)-h^0(K_\Sigma^{1/2}\otimes L_\Sigma(q)^*,\Sigma)\\
&=h^0(K_\Sigma^{1/2}\otimes L_\Sigma(q),\Sigma)-h^1(K_\Sigma^{1/2}\otimes L_\Sigma(q),\Sigma)\\
&=\chi(K_\Sigma^{1/2}\otimes L_\Sigma(q),\Sigma),
\end{split}
\ee
where we have used Serr\'{e} duality
\be
h^0(L)=h^1(K-L).
\ee

}

\item{6D vector multiplet $R_1=(1/2,1/2)\oplus 2(1/2,0)$

The 6D particle contents of $R_1$ is a massless vector in $\mbf{6}$ representation of SO(5,1) and a chiral spinor in $\mbf{4}$ representation. The vector part has no R-charge, and the branching rule of the 6D vector under (\ref{4Dbranch}) is
\be
(\mbf{6},\mbf{1})\rightarrow(\mbf{2},\mbf{2},0,0)+(\mbf{1},\mbf{1},1,0)+(\mbf{1},\mbf{1},-1,0).
\ee
For the fermionic part, we take two 6D Weyl spinors $\chi^1,\chi^2$ in the $(\mbf{4},\mbf{2})$ representation of $SO(5,1)\times SU(2)_R$. They satisfy the symplectic Majorana conditions:
\be
\bsp
\chi^1&=(\chi^2)^C\\
\chi^2&=-(\chi^1)^C.\label{sympMajorana}
\end{split}
\ee
The branching rule of $(\mbf{4},\mbf{2})$ after the dimensional reduction is
\be
\bsp
(\mbf{4},\mbf{2})\rightarrow &(\mbf{2},\mbf{1},\frac{1}{2},1)+(\mbf{1},\mbf{2},-\frac{1}{2},1)\\
&+(\mbf{2},\mbf{1},\frac{1}{2},-1)+(\mbf{1},\mbf{2},-\frac{1}{2},-1).
\end{split}
\ee
After the topological twist, the 4D bosonic spectrum is
\be
(\mbf{2},\mbf{2})_0+(\mbf{1},\mbf{1})_1+(\mbf{1},\mbf{1})_{-1},
\ee
while the fermionic parts are
\be
(\mbf{2},\mbf{1})_0+(\mbf{1},\mbf{2})_{-1}+(\mbf{2},\mbf{1})_1+(\mbf{1},\mbf{2})_0.
\ee
The bosonic field components in representation $R_q$ and $\bar{R}_{-q}$ of $H$ are:
\be
\bsp
(\mbf{2},\mbf{2})_0: &A_\mu(R_q)\in H^0_{\bar\partial}(L_\Sigma(q),\Sigma)\\
                                 &A^c_\mu(\bar{R}_{-q})\in H^0_{\bar\partial}(L_\Sigma(q)^*,\Sigma)\\
(\mbf{1},\mbf{1})_{-1}: &\phi^1(R_q)\in H^0_{\bar\partial}(K_\Sigma\otimes L_\Sigma(q),\Sigma)\\
                                     &\phi^{1,c}(\bar{R}_{-q})\in H^0_{\bar\partial}(K_\Sigma\otimes L_\Sigma(q)^*,\Sigma)\\
(\mbf{1},\mbf{1})_{1}: &\phi^2(R_q)\in \overline{H^0_{\bar\partial}(K_\Sigma\otimes L_\Sigma(q)^*,\Sigma)}\\
                                     &\phi^{2,c}(\bar{R}_{-q})\in \overline{H^0_{\bar\partial}(K_\Sigma\otimes L_\Sigma(q),\Sigma)},\end{split}\label{4Dvec}
\ee
while the fermionic fields are
\be
\bsp
(\mbf{2},\mbf{1})_{0}: &\eta^1(R_q)\in H^0_{\bar\partial}( L_\Sigma(q),\Sigma)\\
                                     &\eta^{1,c}(\bar{R}_{-q})\in H^0_{\bar\partial}(L_\Sigma(q)^*,\Sigma)\\
(\mbf{1},\mbf{2})_{-1}: &\bar\psi^1(R_q)\in H^0_{\bar\partial}(K_\Sigma\otimes L_\Sigma(q),\Sigma)\\
                                     &\bar\psi^{1,c}(\bar{R}_{-q})\in H^0_{\bar\partial}(K_\Sigma\otimes L_\Sigma(q)^*,\Sigma).\\                         
(\mbf{2},\mbf{1})_1: &\psi^2(R_q)\in \overline{H^0_{\bar\partial}(K_\Sigma\otimes L_\Sigma(q)^*)}\\
                                 &\psi^{2,c}(\bar{R}_{-q})\in \overline{H^0_{\bar\partial}(K_\Sigma\otimes L_\Sigma(q))}\\
(\mbf{1},\mbf{2})_0: &\bar\eta^2(R_q)\in \overline{H^0_{\bar\partial}( L_\Sigma(q)^*,\Sigma)}\\
                                 &\bar\eta^{2,c}(\bar{R}_{-q})\in \overline{H^0_{\bar\partial}( L_\Sigma(q),\Sigma)}
\end{split}\label{4Dvecfermion}
\ee
Note that the cohomology groups of $(\mbf{2},\mbf{1})_{0}$ and $(\mbf{1},\mbf{2})_{0}$ are different. Now the fermionic spectrum (\ref{4Dvecfermion}) is apparently CPT invariant. We thus have the following 4D $\mc{N}=1$ supermultiplets: $h^0(L_\Sigma(q),\Sigma)$ copies of vector multiplet $(A_\mu(R_q),\eta(R_q))$, $h^0(L_\Sigma(q)^*,\Sigma)$ copies of vector multiplet $(A^c_\mu(\bar{R}_{-q}),\eta^c(\bar{R}_{-q}))$, $h^0(K_\Sigma\otimes L_\Sigma(q)^*,\Sigma)$ copies of anti-chiral multiplet $(\psi^1(R_q),\phi^1(R_q))$ and $h^0(K_\Sigma\otimes L_\Sigma(q),\Sigma)$ copies of anti-chiral multiplet $(\psi^{1,c}(\bar{R}_{-q}),\phi^{1,c}(\bar{R}_{-q}))$. 

The net number of chiral generations in representation $R_q$ from the vector multiplets is then
\be
\chi^v(R_q)=h^0(L_\Sigma(q),\Sigma)-h^0(L_\Sigma(q)^*,\Sigma),\label{chiv}
\ee
while the net number of chiral generations in representation $R_q$ from the chiral multiplets is
\be
\bsp
\chi^c(R_q)=&-h^0(K_\Sigma\otimes L_\Sigma(q),\Sigma)\\
&+h^0(K_\Sigma\otimes L_\Sigma(q)^*,\Sigma).\label{chic}
\end{split}
\ee
In total, we have
\be
\bsp
\chi(R_q)=&h^0(L_\Sigma(q),\Sigma)-h^0(L_\Sigma(q)^*,\Sigma)\\
&-h^0(K_\Sigma\otimes L_\Sigma(q),\Sigma)\\
&+h^0(K_\Sigma\otimes L_\Sigma(q)^*,\Sigma).\label{chivector}
\end{split}
\ee
Finally, we need to impose the conditions (\ref{sympMajorana}) on the 4D spinors, which is non-trivial to write down with the twists. Effectively, the net degree of freedom and chirality will be reduced to one half of that in (\ref{chivector}).

Note that (\ref{chiv},\ref{chic}) cannot be written as the Euler characteristic of a line bundle. Nonetheless, if $\Sigma=\mb{P}^1$, the dimension of Dolbeault cohomology group
\be
\bsp
&\mathrm{dim}(H^0_{\bar\partial}(K_\Sigma\otimes L_\Sigma(q),\Sigma))\\
&\qquad\qquad\qquad =h^0(K_\Sigma\otimes L_\Sigma(q),\Sigma)\label{vectorP1}
\end{split}
\ee
is always a topological invariant. The reason is that $\mb{P}^1$ and any line bundles on $\mb{P}^1$ do not have any complex structure. 

Note that the 4D complex vector fields have different multiplicities under representation $R_q$ and $\bar{R}_{-q}$. Since $A_\mu(R_q)$ and $A^c_\mu(\bar{R}_{-q})$ are not in the adjoint representation of a Lie group, the counting of on-shell degree of freedom is incorrect if these 4D vector fields are massless. If there is no gauge symmetry associated to these vector fields, the number of on-shell d.o.f. will be 3 instead of 2. 

For this reason, we speculate that the a mass $m_A$ should be generated for these 4D vector fields in presence of the gauge flux. The detail of this physical mechanism is unknown at this point, and we will leave it to future research.

}

\item{6D Rarita-Schwinger multiplet $R_2=(1,1/2)\oplus 2(1,0)$

For higher spin fields, the group decomposition under (\ref{4Dbranch}) is more and more complicated. We will only consider the example of a Rarita-Schwinger multiplet, where the fermionic and bosonic components are under $\overline{\mbf{20}}$ and $\mbf{15}$ representation of SO(5,1) respectively. The branching rules are
\be
\bsp
(\overline{\mbf{20}},\mbf{1})\rightarrow & (\mbf{3},\mbf{2},\frac{1}{2},0)+(\mbf{2},\mbf{3},-\frac{1}{2},0)\\
&+(\mbf{2},\mbf{1},\frac{3}{2},0)+(\mbf{2},\mbf{1},-\frac{1}{2},0)+
(\mbf{1},\mbf{2},\frac{1}{2},0)\\
&+(\mbf{1},\mbf{2},-\frac{3}{2},0)\\
(\mbf{15},\mbf{2})\rightarrow & (\mbf{3},\mbf{1},0,1)+(\mbf{1},\mbf{3},0,1)\\
&+(\mbf{2},\mbf{2},1,1)+(\mbf{2},\mbf{2},-1,1)+(\mbf{1},\mbf{1},0,1)\\
&+(\mbf{3},\mbf{1},0,-1)+(\mbf{1},\mbf{3},0,-1)+(\mbf{2},\mbf{2},1,-1)\\
&+(\mbf{2},\mbf{2},-1,-1)+(\mbf{1},\mbf{1},0,-1)\label{eq:6D_decomp_1}
\end{split}
\ee
After the topological twist, the 4D spectrum is
\be
\bsp
&(\mbf{3},\mbf{2})_{1/2}+(\mbf{2},\mbf{3})_{-1/2}+(\mbf{2},\mbf{1})_{3/2}+(\mbf{2},\mbf{1})_{-1/2}\\
&+(\mbf{1},\mbf{2})_{1/2}+(\mbf{1},\mbf{2})_{-3/2}+(\mbf{3},\mbf{1})_{-1/2}+(\mbf{1},\mbf{3})_{-1/2}\\
&+(\mbf{2},\mbf{2})_{1/2}+(\mbf{2},\mbf{2})_{-3/2}+(\mbf{1},\mbf{1})_{-1/2}+(\mbf{3},\mbf{1})_{1/2}\\
&+(\mbf{1},\mbf{3})_{1/2}+(\mbf{2},\mbf{2})_{3/2}+(\mbf{2},\mbf{2})_{-1/2}+(\mbf{1},\mbf{1})_{1/2}.
\end{split}
\ee
Applying (\ref{negJtop}, \ref{posJtop}), we get the following fermionic spectrum in representation $R_q$:
\be
\bsp
&(\mbf{2},\mbf{3})_{-1/2}: \psi_\alpha^\mu(R_q)\in H^0_{\bar\partial}(K_\Sigma^{1/2}\otimes L_\Sigma(q),\Sigma)\\
&(\mbf{3},\mbf{2})_{1/2}: \bar{\psi}_{\dot\alpha}^\mu(R_q)\in \overline{H^0_{\bar\partial}(K_\Sigma^{1/2}\otimes L_\Sigma(q)^*,\Sigma)}\\
&(\mbf{2},\mbf{1})_{-1/2}: \psi_\alpha(R_q)\in H^0_{\bar\partial}(K_\Sigma^{1/2}\otimes L_\Sigma(q),\Sigma)\\
&(\mbf{1},\mbf{2})_{1/2}: \bar{\psi}_{\dot\alpha}(R_q)\in \overline{H^0_{\bar\partial}(K_\Sigma^{1/2}\otimes L_\Sigma(q)^*,\Sigma)}\\
&(\mbf{2},\mbf{1})_{3/2}: \eta_\alpha(R_q)\in \overline{H^0_{\bar\partial}(K_\Sigma^{3/2}\otimes L_\Sigma(q)^*,\Sigma)}\\
&(\mbf{1},\mbf{2})_{-3/2}: \bar{\eta}_{\dot\alpha}(R_q)\in H^0_{\bar\partial}(K_\Sigma^{3/2}\otimes L_\Sigma(q),\Sigma)
\end{split}
\ee

The net chirality of representation $R_q$ from the 4D Rarita-Schwinger supermultiplet is then
\be
\chi^{RS}(R_q)=\chi(K_\Sigma^{1/2}\otimes L_\Sigma(q),\Sigma).
\ee
The chirality from the 4D chiral multiplet with $J_{top}=\pm 1/2$ and $J_{top}=\pm 3/2$ is
\be
\bsp
\chi^c(R_q)=&\chi(K_\Sigma^{1/2}\otimes L_\Sigma(q),\Sigma)-h^0(K_\Sigma^{3/2}\otimes L_\Sigma(q),\Sigma)\\
&+h^0(K_\Sigma^{3/2}\otimes L_\Sigma(q)^*,\Sigma).
\end{split}
\ee
}

\end{enumerate}

\section{Discussions}
\label{sec:discussion}

In this section, we are going to discuss various caveats and future directions. We shall start with the 6D physics involving the E-string theory and non-flat fiber.

\vspace*{0.1in}
 \noindent
{\bf 6D E-string spectrum}

In this paper, we have obtained an infinite massless tower of M2 brane wrapping modes with various $E_8$ representation and spin, see section~\ref{sec:6Dspectrum}. The lowest representation is a 6D hypermultiplet in representation $\mbf{248}$ of $E_8$, which is consistent with the former results~\cite{Klemm:1996hh}. However, the exact role of these particle modes in the E-string theory is still unclear. It would be interesting to construct an infinite tower of primary operators from these M2 brane wrapping modes, and compare them with bootstrap results~\cite{Chang:2017xmr}. 

Besides the M2 brane wrapping modes, there also exist M5 brane wrapping modes over the whole non-flat fiber $S_{nf}$, which give rise to tensionless string modes in the 6D E-string theory. We currently have no tool to read out the string modes on these tensionless strings, but they would be crucial to fully understand the E-string spectrum.

In this paper, we have only studied one particular resolution of the singular Weierstrass model over $\mb{F}_{11}$. It would be interesting to classify the different Coulomb branches of the gauged E-string theory by studying other different resolutions\footnote{After the first draft of this paper was posted online, several cases of non-flat fiber with different topology was presented in~\cite{Apruzzi:2018nre}}. Such classification of Coulomb branches has already been worked out for a number of gauge field theories coupled with charged  hypermultiplets, such as~\cite{Hayashi:2013lra,Hayashi:2014kca,Esole:2014bka,Esole:2014hya,Braun:2014kla,Esole:2018mqb}.

\vspace*{0.1in}
 \noindent
{\bf Coupling SCFT to gravity}

Using the 6D F-theory on a generic fibration $X_3$ over $\mb{F}_{11}$, we have actually found a way to couple the E-string theory with $E_8$ gauge theory and supergravity. Naturally, the 6D gauge and gravity anomaly cancellation equations should hold as well, see~\cite{Kumar:2010ru}. For example, the 6D gravity cancellation equation
\be
H_{\rm charged}+H_{\rm neutral}-V=273-29T
\ee
 on $X_3$ is satisfied with
 \be
 \bsp
 H_{\rm neutral}&=463\\
 V&=248\\
 T&=1
 \end{split}
 \ee 
and
\be
H_{\rm charged}=29.
\ee
Hence the 6D E-string theory effectively contributes as 29 free charged hypermultiplet when it is coupled with gravity (see also~\cite{Buchmuller:2017wpe,Dierigl:2018nlv}. From this observation, the massless E-string spectrum couples to the $E_8$ gauge field and gravity in a very non-trivial way, which it is interesting to investigate in detail.

\vspace*{0.1in}
 \noindent
{\bf Compactification of strongly coupled matter}

Ultimately, we would like to study the strongly coupled matter fields in 4D from the compactification of 6D E-string theory on $\Sigma$, which couple to the $E_8$ gauge group and gravity. Although we can reduce the 6D E-string operators to 4D, it is unclear whether these 4D fields can be thought as particles with asymptotic states. After the inclusion of gauge flux on $\Sigma$, it is also possible that many of these fields will become massive, see the discussion after (\ref{vectorP1}). We neither know much about the RG flow of 4D gauge theory coupled with strongly coupled matter, and what are the massless degree of freedom in the IR. They will be crucial if one hope to construct standard model matter spectrum.

We shall leave this to future research.

\vspace*{0.1in}
 \noindent
{\bf Other 4D conformal matter}

In this paper, we have only considered the case of E-string theory compactified on a Riemann surface. On a general complex threefold base $B_3$, there are other types of (4,6) curves on an $E_8$ divisor as well. For the class II (4,6) curves, the localized  physical spectrum cannot be computed from the compactification of a 6D SCFT. For these more general cases, we need to directly study the M2 brane wrapping modes on non-flat fiber from Calabi-Yau fourfold geometry, which is an interesting next project to study.

\acknowledgments

We would like to thank Chris Beasley, Chi-Ming Chang, James Halverson, Ling Lin, Cody Long, Pran Nath, Benjamin Sung for helpful discussions. Especially, we would like to thank Sakura Schafer-Nameki and Washington Taylor for comments on the manuscript. We also want to thank the organizers of Tsinghua Summer Workshop in Geometry and Physics 2018 at YMSC, Tsinghua for their hospitality. Y.-N. W. is supported by the ERC Consolidator Grant 682608 “Higgs bundles: Supersymmetric Gauge Theories and Geometry (HIGGSBNDL)”.

\appendix

\section{Computation of intersection numbers on a toric variety}
\label{app:selfint}

In this section, we present the methodology of computing intersection numbers of divisors on a compact toric variety, as mentioned in section~\ref{sec:resolution}. We will consider a 4D toric variety $T_\Sigma$ with toric fan $\Sigma$ in our discussion, although the procedure can be easily generalized to other dimensions as well.

The input data of $T_\Sigma$ consists of the set of 1D toric rays $v_i\in\mb{Z}^4$:
\be
\Sigma(1)=\{v_i=(v_{i,1},v_{i,2},v_{i,3},v_{i,4}) (i=1,\dots,n)\}
\ee
and the set of 4D cones
\be
\Sigma(4)=\{v_i v_j v_k v_l (i\neq j\neq k\neq l)\}.
\ee
We denote the toric divisor corresponding to $v_i$ by $\hat{D}_i$.

With the set of 4D cones, we can easily compute the set of 2D cones
\be
\Sigma(2)=\{v_i v_j|\exists v_i v_j v_k v_l\in\Sigma(4)\}
\ee
and the set of 3D cones
\be
\Sigma(3)=\{v_i v_j v_k|\exists v_i v_j v_k v_l\in\Sigma(4)\}.
\ee
Then for any $v_i v_j v_k v_l\in\Sigma(4)$, their intersection numbers are
\be
\hat{D}_i\cdot \hat{D}_j\cdot \hat{D}_k\cdot \hat{D}_l=\frac{1}{|Vol(v_i v_j v_k v_l)|},\label{quarint}
\ee
where $Vol(v_i v_j v_k v_l)$ is the volume of the 4D cone.

For the other intersection numbers involving self-intersections $D_i^2$, $D_i^3$ or $D_i^4$, we can compute them with the four linear equivalence relations
\be
\bsp
&\sum_{i=1}^{|\Sigma(1)|} v_{i,j}D_i=0\ ,\\
&(j=1,\dots,4).\label{4Dlinequiv}
\end{split}
\ee
We take the product of equation (\ref{4Dlinequiv}) with a product of three toric divsiors $D_i\cdot D_j\cdot D_k$, and get the following different types of equations:
\be
\forall v_k v_l v_m\in\Sigma(3)\ ,\ \sum_{i=1}^{|\Sigma(1)|} v_{i,j}D_i\cdot D_k\cdot D_l\cdot D_m=0\ ,\ (j=1,\dots,4),\label{linequiv1}
\ee
\be
\bsp
&\forall v_k v_l\in\Sigma(2)\ ,\\
&\sum_{i=1}^{|\Sigma(1)|} v_{i,j}D_i\cdot D_k^2\cdot D_l=0= \sum_{i=1}^{|\Sigma(1)|} v_{i,j}D_i\cdot D_k\cdot D_l^2=0\ ,\\
&(j=1,\dots,4),\label{linequiv2}
\end{split}
\ee
\be
\forall v_k\in\Sigma(1)\ ,\ \sum_{i=1}^{|\Sigma(1)|} v_{i,j}D_i\cdot D_k^3=0\ ,\ (j=1,\dots,4).\label{linequiv3}
\ee
We first solve the equations (\ref{linequiv1}) to compute all the intersection numbers in the form of $D_i^2\cdot D_j\cdot D_k$. Then we plug them into (\ref{linequiv2}) to solve all the intersection numbers in the form of $D_i^3 D_j$ and $D_i^2 D_j^2$. Finally, we solve (\ref{linequiv3}) to obtain all the intersection numbers in the form of $D_i^4$.

As a concrete example, let us consider the generic $\mb{P}^{2,3,1}$ bundle over $\mb{P}^2$, with the following set of 1D rays $v_i$:
\be
\bsp
\Sigma(1)&=\{v_i\}\\
&=\{(1,0,-2,-3),(0,1,-2,-3),(-1,-1,-2,-3),\\
&\qquad (0,0,1,0), (0,0,0,1), (0,0,-2,-3)\}
\end{split}
\ee
and the 4D cones:
\be
\bsp
\Sigma(4)&=\{v_1 v_2 v_4 v_5,v_1 v_2 v_4 v_6,v_1 v_2 v_5 v_6,v_1 v_3 v_4 v_5,\\
&\quad\quad v_1 v_3 v_4 v_6,v_1 v_3 v_5 v_6,v_2 v_3 v_4 v_5,\\
&\quad\quad v_2 v_3 v_4 v_6,v_2 v_3 v_5 v_6\}
\end{split}
\ee
We can hence compute the set of 2D and 3D cones:
\be
\bsp
\Sigma(2)=&\{v_1 v_2,v_1 v_3,v_1 v_4,v_1 v_5,v_1 v_6,v_2 v_3,v_2 v_4,\\ 
&\quad v_2 v_5,v_2 v_6,v_3 v_4,v_3 v_5,v_3 v_6, v_4 v_5, v_4 v_6, v_5 v_6\}
\end{split}
\ee
\be
\bsp
\Sigma(3)=&\{v_1 v_2 v_4, v_1 v_2 v_5, v_1 v_2 v_6, v_1 v_3 v_4, v_1 v_3 v_5,\\
&\quad v_1 v_3 v_6, v_1 v_4 v_5, v_1 v_4 v_6, v_1 v_5 v_6, v_2 v_3 v_4, v_2 v_3 v_5,\\
&\quad v_2 v_3 v_6, v_2 v_4 v_5, v_2 v_4 v_6, v_2 v_5 v_6, v_3 v_4 v_5, v_3 v_4 v_6,\\
&\quad v_3 v_5 v_6\}.
\end{split}
\ee
The intersection numbers of the form (\ref{quarint}) are
\be
\bsp
&D_1\cdot D_2\cdot D_4\cdot D_5=D_1\cdot D_3\cdot D_4\cdot D_5\\
&\qquad\qquad\qquad\qquad =D_2\cdot D_3\cdot D_4\cdot D_5=1\ ,\ \\
&D_1\cdot D_2\cdot D_4\cdot D_6=D_1\cdot D_3\cdot D_4\cdot D_6\\
&\qquad\qquad\qquad\qquad =D_2\cdot D_3\cdot D_4\cdot D_6=\frac{1}{3}\ ,\ \\
&D_1\cdot D_2\cdot D_5\cdot D_6=D_1\cdot D_3\cdot D_5\cdot D_6\\
&\qquad\qquad\qquad\qquad =D_2\cdot D_3\cdot D_5\cdot D_6=\frac{1}{2}.
\end{split}
\ee
The linear equivalence relations (\ref{4Dlinequiv}) are
\be
\bsp
D_1&=D_3,\\
D_2&=D_3,\\
D_4&=2(D_1+D_2+D_3+D_6),\\
D_5&=3(D_1+D_2+D_3+D_6),
\end{split}
\ee
which means that we can use $(D_3,D_6)$ as a base of the Picard group of $T_\Sigma$. Then from (\ref{linequiv1},\ref{linequiv2},\ref{linequiv3}), we can solve the intersection numbers
\be
D_3^4=D_3^3 D_6=0\ ,\ D_3^2 D_6^2=\frac{1}{6}\ ,\  D_3 D_6^3=-1\ ,\ D_6^4=\frac{9}{2}.
\ee

We can then compute the triple intersection numbers of the Calabi-Yau threefold hypersurface $X\subset T_\Sigma$ with (\ref{tripint}). Since
\be
-K(T_\Sigma)=18D_3+6D_6,
\ee
we can compute
\bea
D_3^3\cdot (-K(T_\Sigma))&=&0\\
D_3^2\cdot D_6\cdot (-K(T_\Sigma))&=&1\\
D_3\cdot D_6^2\cdot (-K(T_\Sigma))&=&-3\\
D_6^3\cdot (-K(T_\Sigma))&=&9.
\eea

\section{Toric fan of $T_{\rm res}$}
\label{app:toricfan}

In this section, we will list all the 4D rays and cones of the 4D toric ambient space $T_{\rm res}$ constructed in section~\ref{sec:resolution}, which is a blow up of $\mb{P}^{2,3,1}$ bundle over $\mb{F}_{11}$. The 4D rays are
\bea
&&\tilde{v}_1=(1,0,-2,-3)\ ,\ \tilde{v}_2=(0,-1,-2,-3),\nn\\
&&\tilde{v}_3=(-1,-11,-2,-3)\ ,\ \tilde{v}_4=(0,1,-2,-3),\nn\\
&&\tilde{v}_5=(0,0,1,0)\ ,\ \tilde{v}_6=(0,0,0,1)\ ,\ \tilde{v}_7=(0,0,-2,-3),\nn\\
&&\tilde{v}_8=(0,-1,-1,-2)\ ,\ \tilde{v}_9=(0,-1,0,-1),\nn\\
&&\tilde{v}_{10}=(0,-1,-1,-1)\ ,\ \tilde{v}_{11}=(0,-1,0,0),\nn\\
&&\tilde{v}_{12}=(0,-2,-1,-2)\ ,\ \tilde{v}_{13}=(0,-2,-2,-3),\nn\\
&&\tilde{v}_{14}=(0,-2,0,-1)\ ,\ \tilde{v}_{15}=(0,-2,-1,-1),\nn\\
&&\tilde{v}_{16}=(0,-3,-1,-2)\ ,\ \tilde{v}_{17}=(0,-4,-2,-3),\nn\\
&&\tilde{v}_{18}=(0,-3,-1,-1)\ ,\ \tilde{v}_{19}=(0,-4,-1,-2),\nn\\
&&\tilde{v}_{20}=(0,-5,-2,-3)\ ,\ \tilde{v}_{21}=(0,-6,-2,-3).\label{4dtoricrays}
\eea
The 4D cones are (we have omitted $\tilde{v}$)
\bea
&&(1,4,5,6), (1,4,5,7), (1,4,6,7), (3,4,5,6),\nn\\
&&(3,4,5,7), (3,4,6,7), (1,2,5,7), (1,2,6,7),\nn\\
&&(2,3,5,7), (2,3,6,7), (1,2,6,10), (1,6,10,15),\nn\\
&&(1,6,15,18), (1,6,11,18), (1,5,6,11), (1,2,10,13)\,\nn\\
&&(1,2,8,13), (1,2,5,8) , (1,5,8,9), (1,5,9,14),\nn\\
&&(1,5,11,14), (1,8,9,12), (1,12,16,17), (1,9,12,14),\nn\\
&&(1,12,14,16), (1,11,18,19), (1,11,14,19), (1,18,19,21),\nn\\
&&(1,18,20,21), (1,15,18,20), (1,14,16,19), (1,16,19,21),\nn\\
&&(1,16,20,21), (1,10,15,17), (1,15,17,20), (1,16,17,20),\nn\\
&&(1,8,12,13), (1,10,12,13), (1,10,12,17), (3,2,6,10),\nn\\
&&(3,6,10,15), (3,6,15,18), (3,6,11,18), (3,5,6,11),\nn\\
&&(3,2,10,13), (3,2,8,13), (3,2,5,8), (3,5,8,9),\nn\\
&&(3,5,9,14), (3,5,11,14), (3,8,9,12), (3,12,16,17),\nn\\
&&(3,9,12,14), (3,12,14,16), (3,11,18,19), (3,11,14,19),\nn\\
&&(3,18,19,21), (3,18,20,21), (3,15,18,20), (3,14,16,19),\nn\\
&&(3,16,19,21), (3,16,20,21), (3,10,15,17), (3,15,17,20),\nn\\
&&(3,16,17,20), (3,8,12,13), (3,10,12,13), (3,10,12,17)\nn\\
&&
\eea

We list the correspondence between the rays in (\ref{4dtoricrays}) and the divisors mentioned in section~\ref{sec:resolution}:
\bea
&&\tilde{v}_1:v=0\ ,\ \tilde{v}_2:u=0\ ,\ \tilde{v}_5:x=0\ ,\ \tilde{v}_6:y=0\ ,\nn\\
&&\tilde{v}_{13}:E_1\ ,\ \tilde{v}_{10}:E_2\ ,\ \tilde{v}_{17}:E_3\ ,\ \tilde{v}_{20}:E_4\ ,\ \tilde{v}_{21}:E_5\ ,\nn\\
&&\tilde{v}_{19}:E_6\ ,\ \tilde{v}_{14}:E_7\ ,\ \tilde{v}_{21}:E_8.
\eea

\section{Group theory results}
\label{app:group}

We summarize the relevant group theory facts in this section~\cite{Feger:2012bs,Yamatsu:2015npn}.

\subsection{$E_8$ representations}

\begin{figure}
\centering
\includegraphics[height=2cm]{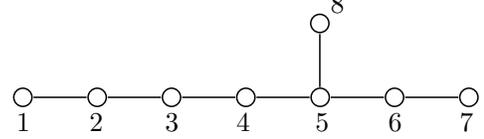}
\caption[E]{\footnotesize The labeling of 8 Dynkin nodes of $E_8$.}\label{f:E8}
\end{figure}

We label the 8 Dynkin nodes of $E_8$ in figure~\ref{f:E8}. Then the irreducible representations of $E_8$ are labelled by their highest weight $(a_1,a_2,a_3,a_4,a_5,a_6,a_7,a_8)$. We list a few of the lowest dimensional representations in table~\ref{t:E8rep}.

\begin{table}
\centering
\begin{tabular}{|c|c|}
\hline
\hline
Highest weight & Representation\\
\hline
(0,0,0,0,0,0,0,0) & $\mbf{1}$\\
(1,0,0,0,0,0,0,0) & $\mbf{248}$\\
(0,0,0,0,0,0,1,0) & $\mbf{3875}$\\
(2,0,0,0,0,0,0,0) & $\mbf{27000}$\\
(0,1,0,0,0,0,0,0) & $\mbf{30380}$\\
(0,0,0,0,0,0,0,1) & $\mbf{147250}$\\
(1,0,0,0,0,0,1,0) & $\mbf{779247}$\\
(3,0,0,0,0,0,0,0) & $\mbf{1763125}$\\
(0,0,1,0,0,0,0,0) & $\mbf{2450240}$\\
(1,1,0,0,0,0,0,0) & $\mbf{4096000}$\\
(0,0,0,0,0,0,2,0) & $\mbf{4881384}$\\
(0,0,0,0,0,1,0,0) & $\mbf{6696000}$\\
\hline
\end{tabular}
\caption[E]{\footnotesize The list of the lowest dimensional representations of $E_8$.}\label{t:E8rep}
\end{table}

\bibliography{ref_tian}

\end{document}